\documentclass[a4paper,fleqn,usenatbib]{mnras}

\usepackage[T1]{fontenc}
\usepackage{ae,aecompl}

\usepackage{graphicx}	
\usepackage{amsmath}	
\usepackage{amssymb}	
\usepackage{pdflscape}	

\newcommand{\Lnu}{\ifmmode L_{\rm \nu} \else $L_{\rm \nu}$\fi}
\newcommand{\Llambda}{\ifmmode L_{\rm \lambda} \else $L_{\rm \lambda}$\fi}
\newcommand{\LA}{\ifmmode L_{\rm 5100} \else $L_{\rm 5100}$\fi}
\newcommand{\Lg}{\ifmmode L_{\rm g(r)} \else $L_{\rm g(r)}$\fi}
\newcommand{\Ln}{\ifmmode L_{\rm n(r)} \else $L_{\rm n(r)}$\fi}
\newcommand{\Lo}{\ifmmode L_{\rm 0} \else $L_{\rm 0}$\fi}
\newcommand{\Ro}{\ifmmode R_{\rm 0} \else $R_{\rm 0}$\fi}
\newcommand{\Rhalf}{\ifmmode R_{\rm 1/2} \else $R_{\rm 1/2}$\fi}
\newcommand{\Rg}{\ifmmode R_{\rm g} \else $R_{\rm g}$\fi}
\newcommand{\Rms}{\ifmmode R_{\rm ms} \else $R_{\rm ms}$\fi}
\newcommand{\Rin}{\ifmmode R_{\rm in} \else $R_{\rm in}$\fi}
\newcommand{\Rout}{\ifmmode R_{\rm out} \else $R_{\rm out}$\fi}
\newcommand{\RISCO}{\ifmmode R_{\rm in} \else $R_{\rm in}$\fi}
\newcommand{\ro}{\ifmmode r_{\rm 0} \else $r_{\rm 0}$\fi}
\newcommand{\rhalf}{\ifmmode r_{\rm 1/2} \else $r_{\rm 1/2}$\fi}
\newcommand{\rg}{\ifmmode r_{\rm g} \else $r_{\rm g}$\fi}
\newcommand{\rms}{\ifmmode r_{\rm ms} \else $r_{\rm ms}$\fi}
\newcommand{\rin}{\ifmmode r_{\rm in} \else $r_{\rm in}$\fi}
\newcommand{\rout}{\ifmmode r_{\rm out} \else $r_{\rm out}$\fi}
\newcommand{\rISCO}{\ifmmode r_{\rm in} \else $r_{\rm in}$\fi}
\newcommand{\Fnu}{\ifmmode F_{\nu} \else $F_{\nu}$\fi}
\newcommand{\Flambda}{\ifmmode F_{\lambda} \else $F_{\lambda}$\fi}
\newcommand{\Mdot}{\ifmmode \dot{M} \else $\dot{M}$\fi}
\newcommand{\mdot}{\ifmmode \dot{m} \else $\dot{m}$\fi}
\newcommand{\Mrdot}{\ifmmode \dot{M}\left(r \right) \else $\dot{M}\left(r \right)$\fi}
\newcommand{\MRdot}{\ifmmode \dot{M}\left(R \right) \else $\dot{M}\left(R \right)$\fi}
\newcommand{\Mrindot}{\ifmmode \dot{M}\left(r_{ISCO} \right) \else $\dot{M}\left(r_{ISCO} \right)$\fi}
\newcommand{\Mroutdot}{\ifmmode \dot{M}\left(r_{out} \right) \else $\dot{M}\left(r_{out} \right)$\fi}
\newcommand{\MBHdot}{\ifmmode \dot{M}_{BH} \else $\dot{M}_{BH}$\fi}
\newcommand{\MBHdotexpct}{\ifmmode \dot{M}_{BHexpected} \else $\dot{M}_{BHexpected}$\fi}
\newcommand{\Medddot}{\ifmmode \dot{M}_{edd} \else $\dot{M}_{edd}$\fi}
\newcommand{\Moutdot}{\ifmmode \dot{M}_{out} \else $\dot{M}_{out}$\fi}
\newcommand{\Mindot}{\ifmmode \dot{M}_{in} \else $\dot{M}_{in}$\fi}
\newcommand{\Mwinddot}{\ifmmode \dot{M}_{wind} \else $\dot{M}_{wind}$\fi}
\newcommand{\MBH}{\ifmmode M_{\rm BH} \else $M_{\rm BH}$\fi}
\newcommand{\mbh}{\ifmmode M_{\rm BH} \else $M_{\rm BH}$\fi}
\newcommand{\Mexp}{\ifmmode M_{\rm 8} \else $M_{\rm 8}$\fi}

\newcommand{\Msun}{\ifmmode M_{\odot} \else $M_{\odot}$\fi}
\newcommand{\msun}{\ifmmode M_{\odot} \else $M_{\odot}$\fi}
\newcommand{\Msunyr}{\ifmmode M_{\odot}\, {\rm yr}^{-1} \else $M_{\odot}\, {\rm yr}^{-1}$}
\newcommand{\msunyr}{\ifmmode M_{\odot}/yr \else $M_{\odot}/yr$\fi}
\newcommand{\avisc}{\ifmmode \alpha_{visc} \else $\alpha_{visc}$\fi}

\newcommand{\Halpha}{\ifmmode {\rm H}\alpha \else H$\alpha$\fi}
\newcommand{\Hbeta}{\ifmmode {\rm H}\beta \else H$\beta$\fi}
\newcommand{\hb}{\ifmmode {\rm H}\beta \else H$\beta$\fi}
\newcommand{\MgII}{\ifmmode {\rm Mg}\,\textsc{ii}\,\lambda2798 \else Mg\,{\sc ii}\,$\lambda2798$\fi}
\newcommand{\mgii}{\ifmmode {\rm Mg}{\textsc{ii}} \else Mg\,{\sc ii}\fi}
\newcommand{\CIV}{\ifmmode {\rm C}\,\textsc{iv}\,\lambda1549 \else C\,{\sc iv}\,$\lambda1549$\fi}
\newcommand{\civ}{\ifmmode {\rm C}\,\textsc{iv} \else C\,{\sc iv}\fi}

\newcommand{\oi}{\ifmmode \left[{\rm O}\,\textsc{i}\right] \else [O\,{\sc i}]\fi}
\newcommand{\OI}{\ifmmode \left[{\rm O}\,\textsc{i}\right]\,\lambda6300 \else [O\,{\sc i}]$\,\lambda6300$ \fi}

\newcommand{\oii}{\ifmmode \left[{\rm O}\,\textsc{ii}\right] \else [O\,{\sc ii}]\fi}
\newcommand{\OII}{\ifmmode \left[{\rm O}\,\textsc{ii}\right]\,\lambda3727 \else [O\,{\sc ii}]\,$\lambda3727$ \fi}
\newcommand{\oiii}{\ifmmode \left[{\rm O}\,\textsc{iii}\right] \else [O\,{\sc iii}]\fi}
\newcommand{\OIII}{\ifmmode \left[{\rm O}\,\textsc{iii}\right]\,\lambda5007 \else [O\,{\sc iii}]\,$\lambda5007$\fi}

\newcommand{\ld}{\ifmmode {\rm lt-days} \else lt-days \fi}

\newcommand{\ergs}{\ifmmode {\rm erg\,s}^{-1} \else erg\,s$^{-1}$ \fi}
\newcommand{\ergcms}{\ifmmode {\rm erg\,cm}^{-2}\,{\rm s}^{-1} \else erg\,cm$^{-2}$\,s$^{-1}$\fi}
\newcommand{\ergcmsA}{\ifmmode{\rm erg}\, {\rm cm}^{-2}\,{\rm s}^{-1}\,{\rm\AA}^{-1} \else erg\, cm$^{-2}$\, s$^{-1}$\, \AA$^{-1}$\fi}
\newcommand{\ergcmsHz}{\ifmmode{\rm erg\,cm}^{-2}\,{\rm s}^{-1}\,{\rm Hz}^{-1} \else erg\,cm$^{-2}$\,s$^{-1}$\,Hz$^{-1}$\fi}
\newcommand{\phcms}{\ifmmode {\rm ph\,cm}^{-2}\,{\rm s}^{-1} \else ,ph\,cm$^{-2}$\,s$^{-1}$\fi}
\newcommand{\phcmsA}{\ifmmode {\rm ph\,cm}^{-2}\,{\rm s}^{-1}\,{\rm\AA}^{-1} \else ph\,cm$^{-2}$\,s$^{-1}$\,\AA$^{-1}$\fi}

\newcommand{\Lsun}{\ifmmode L_{\odot} \else $L_{\odot}$\fi}
\newcommand{\auvo}{\ifmmode \alpha_{\nu,{\rm UVO}} \else $\alpha_{\nu,{\rm UVO}}$\fi}
\newcommand{\Luv}{\ifmmode L_{1450} \else $L_{1450}$\fi}
\newcommand{\Lop}{\ifmmode L_{5100} \else $L_{5100}$\fi}
\newcommand{\Lthree}{\ifmmode L_{3000} \else $L_{3000}$\fi}
\newcommand{\lLthree}{\ifmmode \log\left(\Lthree/\ergs\right) \else $\log\left(\Lthree/\ergs\right)$\fi}
\newcommand{\lledd}{\ifmmode L/L_{\rm Edd} \else $L/L_{\rm Edd}$\fi}
\newcommand{\Ledd}{\ifmmode L/L_{\rm Edd} \else $L/L_{\rm Edd}$\fi}
\newcommand{\lamLlam}{\ifmmode \lambda L_{\lambda} \else $\lambda L_{\lambda}$\fi}
\newcommand{\Lbol}{\ifmmode L_{\rm bol} \else $L_{\rm bol}$\fi}
\newcommand{\lLbol}{\ifmmode \log\left(\Lbol/\ergs\right) \else $\log\left(\Lbol/\ergs\right)$\fi}

\newcommand{\Fthree}{\ifmmode F_{3000} \else $F_{3000}$\fi}

\title[AGN Accretion Discs and Black Hole Spin]
{Active galactic nuclei at $z\sim 1.5$: III. Accretion discs and Black Hole Spin}

\author[D. M. Capellupo et al.]
{D. M. Capellupo$^{1,2}$
\thanks{E-mail:danielc@physics.mcgill.ca (DMC)},
H. Netzer$^{1}$, P. Lira$^{3}$, B. Trakhtenbrot$^{4}$
\thanks{Zwicky Postdoctoral Fellow},
J. Mej\'{i}a-Restrepo$^{3}$
\\
$^{1}$School of Physics and Astronomy, Tel Aviv University, Tel Aviv 69978, Israel\\
$^{2}$Department of Physics, McGill University, Montreal, Quebec, H3A 2T8, Canada\\
$^{3}$Departamento de Astronom\'{i}a, Universidad de Chile, Camino del Observatorio 1515, Santiago, Chile\\
$^{4}$Institute for Astronomy, Dept. of Physics, ETH Zurich, Wolfgang-Pauli-Strasse 27, CH-8093 Zurich, Switzerland
}

\date{Accepted XXX. Received YYY; in original form ZZZ}

\pubyear{2015}

\begin{document}
\label{firstpage}
\pagerange{\pageref{firstpage}--\pageref{lastpage}}
\maketitle

\begin{abstract}
This is the third paper in a series describing the spectroscopic properties of
a sample of 39 AGN at $z \sim 1.5$, selected to cover a large range in black
hole mass (\mbh) and Eddington ratio (\lledd). In this paper, we continue the
analysis of the VLT/X-shooter observations of our sample with the addition
of 9 new sources.
We use an improved Bayesian procedure, which takes into account intrinsic
reddening, and improved \mbh\ estimates, to fit thin accretion
disc (AD) models to the observed spectra and constrain the spin parameter
($a_*$) of the central black holes.
We can fit 37 out of 39 AGN with the thin AD model, and for those with
satisfactory fits, we obtain constraints on the spin parameter of the BHs, with
the constraints becoming generally less well defined with decreasing BH mass.
Our spin parameter estimates range from $\sim$$-$0.6 to maximum spin for our
sample, and our results are consistent with the ``spin-up'' scenario of BH spin
evolution. We also discuss how the results of our analysis vary with the
inclusion of \emph{non-simultaneous} GALEX photometry in our thin AD fitting.
Simultaneous spectra covering the rest-frame optical through far-UV are
necessary to definitively test the thin AD theory and obtain the best
constraints on the spin parameter.
\end{abstract}




\section{Introduction}
\label{intro}

The dominant source of optical-UV emission in active galactic nuclei (AGN) is
likely an accretion flow surrounding a central super-massive black hole (SMBH).
For most cases, it is believed that this accretion flow takes the form of an
optically thick, geometrically thin accretion disc (thin AD), as described in
\citet{Shakura73}.
The physics of an actively accreting BH is governed by three key parameters,
namely its mass (\mbh), spin (defined using the dimensionless parameter $a_*$),
and accretion rate (\Mdot). These
parameters are intimately connected to the nature of the accretion flow around
the BH, and AGN with very large accretion rates are believed to have optically
thick, geometrically thick accretion discs
\citep[``slim'' ADs;][and references therein]{Abramowicz88,Ohsuga11,Netzer13}.

\defcitealias{Capellupo15}{Paper~I}
There are several `standard' models in the literature that predict the emitted
SED of thin ADs, based on the general ideas in \citet{Shakura73} and with
various improvements, including general relativistic (GR) corrections,
radiative transfer in the disc atmosphere, and disc winds
\citep[e.g.][]{Hubeny01,Davis11,Done12,Slone12}. As described in
\citet{Koratkar99} and \citet{Davis11}, as well as in our previous paper,
\citet{Capellupo15} (hereafter, \citetalias{Capellupo15}), early attempts to
fit such thin AD models to observed AGN spectra have generally found that the
theoretical SEDs are significantly bluer than those observed. However, these
studies were likely affected by relatively narrow wavelength coverage, by
potential variability between different observations taken by different
instruments, and/or stellar light contamination at long wavelengths.

Furthermore, while estimates of \mbh\ and \Mdot\ (or the Eddington ratio,
\lledd) have been obtained for many active SMBHs, the spin parameters are
largely unknown. Up until recently, spin measurements have been limited to
X-ray observations of relatively nearby AGN that are able to probe the
innermost regions of the AD. Specifically, high-quality X-ray observations are
required to model the profile of the relativistic 6.4 keV K$\alpha$ line, and
such measurements have been performed for only a handful of AGN at low redshift
\citep[][and references therein]{Fabian00,Brenneman13,Risaliti13,Reis14,Reynolds14}.
The highest redshift AGN with such a measurement so far is at z $\sim$ 0.6, and
this was possible only because it is lensed \citep{Reis14}. A further downside
to this approach is that these measurements cannot distinguish between negative
spin and spin of 0 because the changes are too small in the broad 6.4 keV line
profile. Therefore, a method that is sensitive to the full range of spin
parameters ($-1 \le a_* \le 1$) and can be applied to AGN at larger redshifts
is necessary.

In \citetalias{Capellupo15}, we introduced a new sample of AGN, observed with
a unique instrument, X-shooter, at the VLT \citep{Vernet11}. This sample was
selected based on the BH mass and
the Eddington ratio (\lledd), two of the three fundamental
properties of active BHs. Nothing was known about the spin of
this sample at the time the sample was selected.

Our sample was selected in a narrow redshift range centered around
$z \simeq 1.55$. This redshift was selected so that the four strongest broad
emission lines (BELs; H$\alpha$, H$\beta$, MgII 2800\AA, and CIV 1549\AA) would
fall within the observed spectral range of the X-shooter instrument. This is
important for addressing the physics of BELs and the estimation of \mbh\ based
on these BELs.
Using the X-shooter instrument avoids
the problem of line and continuum variations that arises when observing
individual BELs at different times and with different instruments.
The results of this part of the project are described in
Mej\'{i}a-Restrepo et al. (hereafter, Paper II).

Our work in \citetalias{Capellupo15} showed that with wide, single-epoch
wavelength coverage of the SEDs, the thin AD theory is indeed consistent with
the data for at least 25 out of the 30 AGN we studied, in contrast with many of
the earlier works on AGN SED fitting. Futhermore, we were able to constrain the
spin parameter for those sources with satisfactory thin AD fits to the SEDs.

In the current work, we improve and expand upon the work in
\citetalias{Capellupo15} in three ways. First, we add an additional 9 AGN to
the sample to fill a section of the \mbh$-$\lledd\ plane missing in
\citetalias{Capellupo15}, namely fainter AGN with a combination of smaller
\mbh\ and lower \lledd. Second, we improve our Bayesian AGN SED fitting
procedure by including improved \mbh\ estimates from Paper II and, instead of
applying intrinsic reddening only to those AGN that could not otherwise be fit
with a thin AD SED, as we did in \citetalias{Capellupo15}, we now include an
intrinsic reddening correction in our Bayesian fitting procedure for all
sources.
Third, we investigate the inclusion of archival photometry from GALEX, in
order to extend our wavelength coverage further into the UV.
This allows us to cover a larger portion of the AGN SED that is dominated by
radiation from the AGN accretion disc. Although, this analysis is hampered
by potential variability between the non-simultaneous GALEX and X-shooter
observations.

We summarize the sample selection, observations, and data reduction in
Section~\ref{sec:data}. In Section~\ref{sec:fit}, we describe the thin AD model
we use, our procedure for fitting the model to the data, and the results of
fitting both the X-shooter spectra alone and the combined X-shooter+GALEX SEDs.
In Section~\ref{sec:discuss}, we discuss the implications of our results on the
nature of AGN accretion discs and our understanding of AGN BH spin evolution.
Throughout this work, we assume a $\Lambda$CDM cosmological model with
$\Omega_{\Lambda}=0.7$, $\Omega_{m}=0.3$, and
$H_{0}=70\, {\rm km\, s^{-1}} \, {\rm Mpc}^{-1}$.

\section{Sample Observations and Data Reduction}
\label{sec:data}

\subsection{X-shooter}
\label{sec:xsh}

\begin{figure}
 \centering
 \includegraphics[width=80mm]{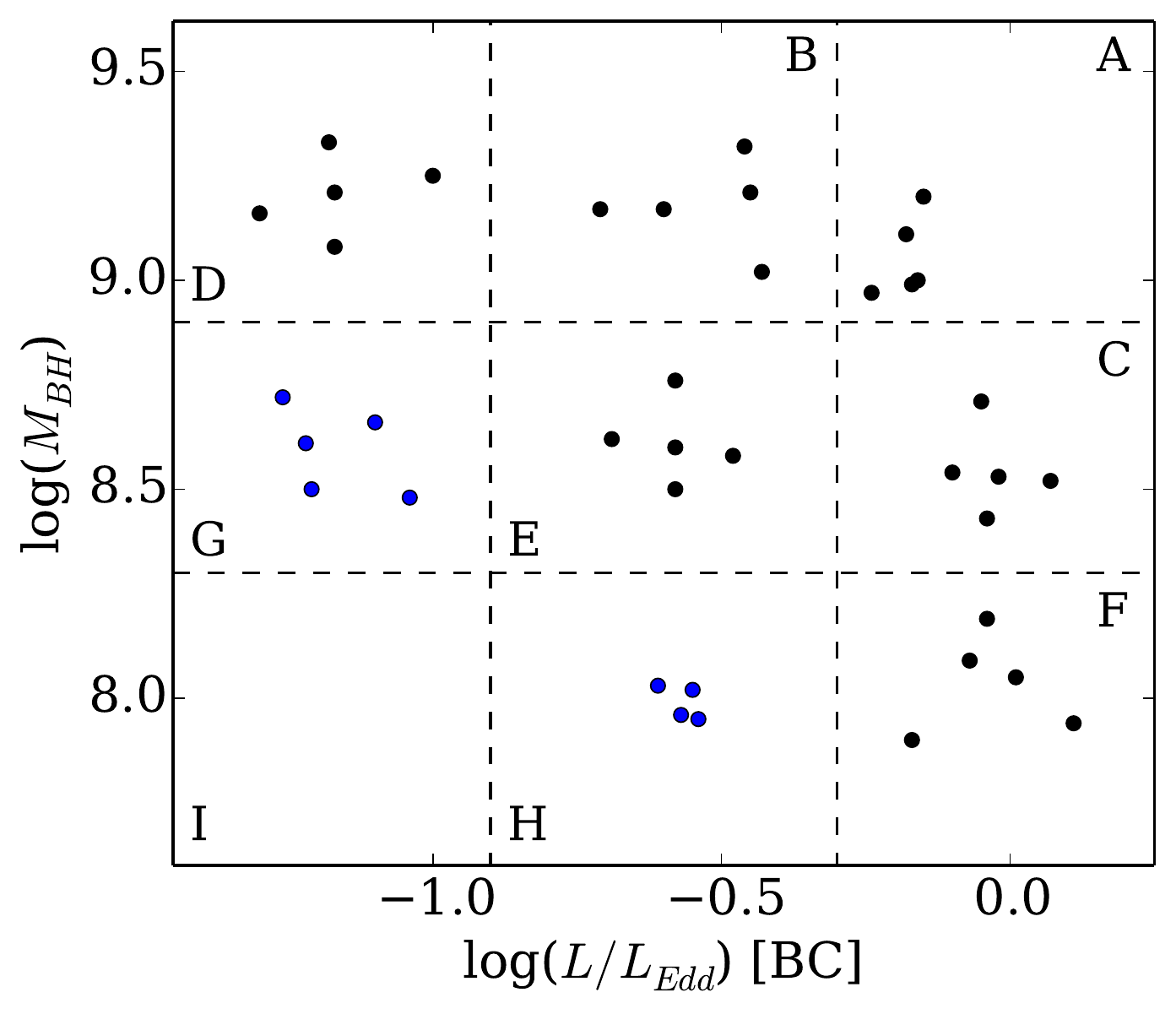}
 \caption{Our sample selection plotted on the $M_{BH}$--$L/L_{Edd}$ plane,
    using the measured values based on SDSS and 2SLAQ spectra and
    \citet{McLure04}. Black points are the original 30 and blue points are the
    nine new sources.}
 \label{fig:mbh_ledd}
\end{figure}

In this work, we use a sample of AGN selected from the seventh data release of
the SDSS \citep{Abazajian09}, as described in \citetalias{Capellupo15}, and from 2SLAQ
\citep{Croom09}. To summarize, our sample
was selected to cover the widest possible range in $M_{BH}$ and $L/L_{Edd}$,
within a narrow redshift range, z $\simeq$ 1.45 -- 1.65. For the purpose of
selecting the sample only, we use measurements of the \mgii\ emission line in
the SDSS (for the original 30 sources described in \citetalias{Capellupo15}) and 2SLAQ (for the
new nine sources presented here) spectra, along with a standard bolometric
correction (BC) factor and relations given in \citet{McLure04}, to estimate
$M_{BH}$ and $L/L_{Edd}$. We divide the known $M_{BH}$--$L/L_{Edd}$ plane into
9 bins, and we select five objects per bin (Fig. \ref{fig:mbh_ledd}). We have
currently observed 39 AGN, in bins A$-$H, with $M_{BH}$ ranging from
$\sim$$9 \times 10^{7}$ to $4 \times 10^{9}$ \Msun\ and $L/L_{Edd}$ from
$\sim$0.04 to 0.7.

\begin{table}
  \caption{Summary of observations and data reduction.}
  \begin{tabular}{clcc}
    \hline
    Name & Dates observed & $A_V^{(a)}$ & Notes \\
    \hline
    J0042$+$0008  & 2013 October 24  & 0.02 & c \\
                  & 2013 October 31  &  & \\
    J1021$-$0027  & 2014 February 23 & 0.05 & b \\
                  & 2014 February 26 &  &  \\
                  & 2014 February 27 &  &  \\
                  & 2014 April 24    &  &  \\
                  & 2014 April 24    &  &  \\
                  & 2014 April 27    &  &  \\
    J0038$-$0019  & 2013 November 03 & 0.02 & c \\
                  & 2013 November 03 &  &  \\
                  & 2013 November 04 &  &  \\
                  & 2013 November 08 &  &  \\
    J0912$-$0040  & 2013 December 31 & 0.03 & c \\
                  & 2014 January 30  &  &  \\
    J1048$-$0019  & 2014 April 27    & 0.04 & c \\
                  & 2015 January 27  &  &  \\
                  & 2015 January 27  &  &  \\
                  & 2015 January 27  &  &  \\
    J1045$-$0047  & 2014 March 08    & 0.04 &  \\
                  & 2014 March 08    &  &  \\
                  & 2014 April 23    &  &  \\
                  & 2014 April 23    &  &  \\
                  & 2014 April 24    &  &  \\
    J0042$-$0011  & 2013 November 04 & 0.02 & c \\
                  & 2013 November 08 &  &  \\
                  & 2014 July 28     &  &  \\
                  & 2014 July 28     &  &  \\
                  & 2014 July 29     &  &  \\
                  & 2014 July 29     &  &  \\
    J1046$+$0025  & 2014 February 24 & 0.04 & c \\
                  & 2014 February 26 &  &  \\
                  & 2014 February 26 &  &  \\
                  & 2014 February 27 &  &  \\
                  & 2014 March 01    &  &  \\
                  & 2014 March 01    &  &  \\
    J0930$-$0018  & 2014 February 04 & 0.03 & c \\
                  & 2014 February 22 &  &  \\
                  & 2014 February 22 &  &  \\
                  & 2014 February 23 &  &  \\
    \hline
  \end{tabular}

  $(a)$ {Galactic extinction.}

  $(b)$ {BALQSO}

  $(c)$ {Requires host galaxy subtraction.}
  \label{tab:data}
\end{table}

The X-shooter instrument at the VLT provides spectra with continuous wavelength
coverage from $\sim$3000 to 25 000 \AA, by simultaneously observing three
wavelength regions, the UV-blue (UVB), visible (VIS), and near-infrared
(NIR; \citealt{Vernet11}). The instrumental set-up for the 9 new sources
presented in this paper (ESO program 092.B-0613) is the same as for the
original 30 (\citetalias{Capellupo15}; ESO program 088.B-1034). We observe with
the widest available slit widths, 1.2 to 1.6 arcsec, giving a resolving power
of 3300 to 5400, depending on the arm. Table \ref{tab:data} lists the nine new
objects in our sample and the dates of observation.

The spectra were reduced using the ESO Reflex environment \citep{Freudling13}
and version 2.5.2 of the ESO X-shooter pipeline, in nodding mode
\citep{Modigliani10}. The pipeline subtracts the detector bias and dark
current, rectifies and calibrates the wavelength scale of the spectra, and uses
an observed spectroscopic standard star spectrum to calculate an absolute
flux-calibrated spectrum. In general, the standard star is observed the same
night as the science target.

With the pipeline-calibrated result, we then corrected the spectra for telluric
absorption within the VIS arm spectrum, using a telluric standard star
observation at a similar airmass as the AGN observation taken either right
before or right after the AGN observation. In the case of the wavelength region
$\sim$8950$-$9800\AA, we used a model telluric spectrum instead of a standard
star observation. In the NIR arm, where there is more significant telluric
absorption, we simply remove the regions of the spectrum most affected by this
absorption.

Finally, we use the \citet{Schlegel98} maps and \citet{Cardelli89} extinction
law to correct the spectra for Galactic extinction. Table \ref{tab:data} lists
the values of $A_V$ due to the Galaxy for the nine new targets.

\begin{figure*}
 \centering
 \includegraphics[width=160mm]{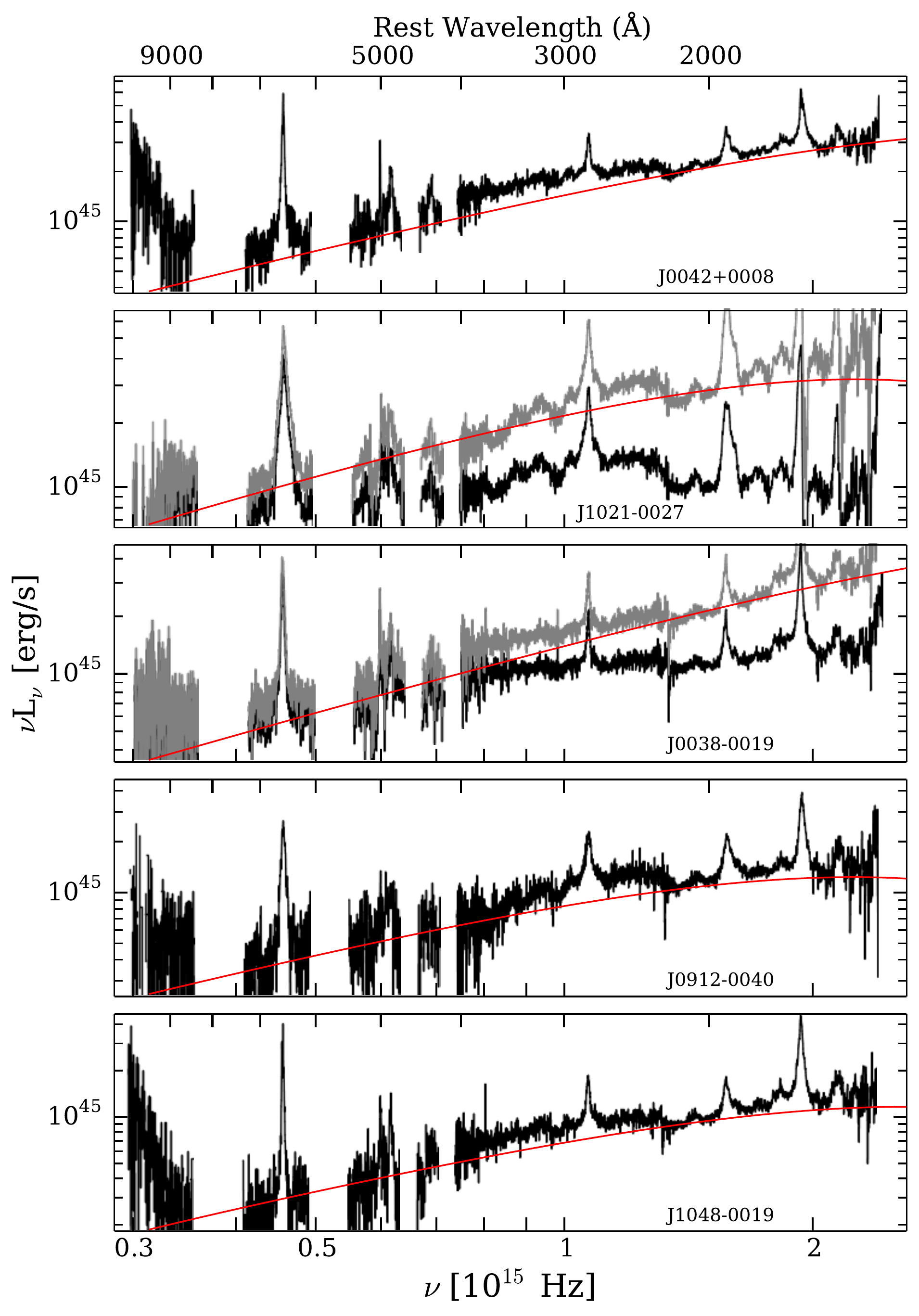}
 \caption{Spectra of the nine new X-shooter sources with the best-fit thin AD
   models (red curves) over-plotted.
   For those objects whose best model fit required an intrinsic reddening
   correction, we plot the dereddened spectrum in gray.
   Seven of the 9 spectra were corrected for host galaxy contribution
   before fitting. The objects are ordered by source luminosity, as determined
   from $\lambda L_{\lambda}(3000)$\AA.
   }
 \label{fig:sp1}
\end{figure*}
\addtocounter{figure}{-1}
\begin{figure*}
 \centering
 \includegraphics[width=160mm]{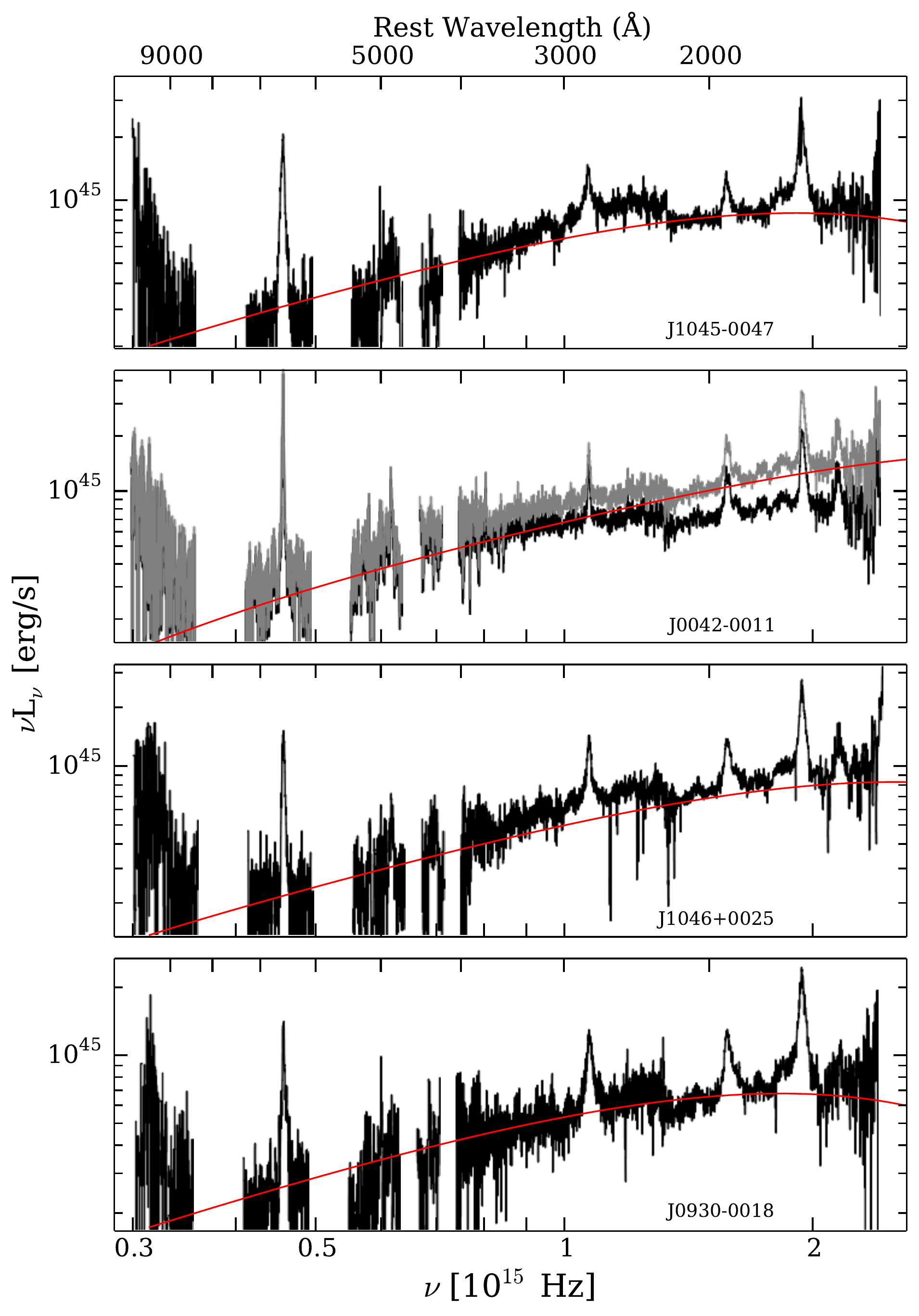}
 \caption{continued...}
\end{figure*}

Fig. \ref{fig:sp1} shows the full X-shooter spectra of the nine new sources.
All sources are corrected for Galactic extinction, and some have been corrected
for host galaxy contamination, as described in Section \ref{sec:thinad}. The
spectra are ordered by source luminosity as determined from
$\lambda L_{\lambda}(3000)$\AA. For consistency, the sources are ordered in
this same way in Table \ref{tab:data}.

\subsection{GALEX}
\label{sec:galex}

To increase our wavelength coverage, we incorporate measurements from the sixth
and seventh data release of GALEX. The GALEX mission has surveyed the sky in
two UV bands. The far-UV filter has a bandwidth of 1344--1786 \AA, with an
effective wavelength of 1538.6 \AA, and the near-UV filter has a bandwidth of
1771--2831 \AA, with an effective wavelength of 2315.7 \AA\
\citep{Morrissey07}. This corresponds to rest wavelengths of $\sim$600 and
900 \AA\ for our sample. The GALEX catalog contains photometric measurements of
38 out of 39 of the AGN in our sample in the NUV band and 20 in the FUV band.
We have up to five epochs of GALEX photometry per source, taken anywhere from
September 2003 to February 2012. The first X-shooter observations from
\citetalias{Capellupo15} began in October 2011, whereas 75\% of our sample only
have GALEX data from before 2010. As detailed below, time variability is
evident in many of these observations, so we consider all epochs here.

The GALEX magnitudes range from 17.7 to 23.5, and most of the errors range from
0.02 to 0.3 mag, with a few measurements having errors as high as 0.6 mag. We
also corrected the GALEX measurements for Galactic extinction, using the same
extinction maps and extinction law as for the X-shooter spectra.

\section{Fitting Accretion Disc Models}
\label{sec:fit}

\subsection{Standard thin AD models}
\label{sec:thinad}

As described in \citetalias{Capellupo15}, most current AD models are based on the blackbody thin
disc model of \citet{Shakura73}, with two significant improvements: the
inclusion of general relativity (GR) terms and the improvement of the radiative
transfer in the disc atmosphere \citep[e.g.][]{Hubeny01,Davis11}. In the
current paper, we continue to use the numerical code presented in
\citet{Slone12} to calculate thin AD spectra, with a viscosity parameter
($\alpha$) of 0.1.

Before calculating thin AD models, we calculate both \mbh\ and \Mdot\ (the
accretion rate in \Msunyr) directly
from the observed spectrum. A major improvement relative to
\citetalias{Capellupo15} is the use of new \mbh\ measurements based on the
comparison of four strong emission lines in our own sample $-$ H$\alpha$,
H$\beta$, \mgii, and \civ\ (Paper II).
The main results of Paper II are:
1) H$\alpha$, H$\beta$, and \mgii\ give consistent estimates of \mbh, albeit
with a normalization which is somewhat different from the one used in
\citetalias{Capellupo15} (based on the \citealt{Trakhtenbrot12} calibration of
the \mgii\ method). The \mgii-based estimates are less reliable for broad
absorption line AGNs and for sources where FWHM(\mgii)$>$FWHM(H$\beta$).
2) The \civ\ line by itself does not provide reliable BH mass estimates for
many of the sources.
3) New estimates of \mbh\ that are based on
the FWHM of \mgii\ are larger than the estimates used in
\citetalias{Capellupo15} by $\sim$0.16 dex, with a scatter of 0.20 dex. All
calculations and model fitting presented in this paper use the new mass
measurements.

The method for measuring \Mdot, in units of \Msunyr, is the
same as in \citetalias{Capellupo15}, and is based on the properties
of thin ADs \citep{Collin02,Davis11} and the fact that thin AD SEDs can be
described by a canonical power law of the form $L_{\nu} \propto \nu^{1/3}$ at
long enough wavelengths. Using the measured \mbh\ and equation 1 from \citetalias{Capellupo15},
we can determine the mass accretion rate directly from the monochromatic
luminosity in a wavelength region along this power law portion of the SED. The
one additional unknown is the inclination of the disc with respect to our line
of sight.

The nine new sources presented here are fainter than the 30 sources presented
in \citetalias{Capellupo15}, and therefore, they are more susceptible to host
galaxy contamination at longer wavelengths, including the wavelength region
used for measuring the accretion rate. We therefore have to subtract the host
galaxy emission in order to more accurately measure the AGN SED.

We determine which objects require a host galaxy subtraction based on the
rest-wavelength equivalent width (EW) of the H$\alpha$ emission line. The EW
of the Balmer lines is not affected by the Baldwin effect, and the H$\alpha$
line intensity is a reliable bolometric luminosity indicator \citep{Stern12}.
We first look at the EW distribution of the brightest 28 AGN in the sample,
whose luminosity at 5100\AA\ is high enough that host contamination is small
enough to safely be neglected \citep{Shen11}. We then compare the EW
distribution for the 11 faintest AGN in the sample to the distribution for the
brighter AGN, and we find most of the faint AGN have EW smaller than the median
EW of the brighter AGN (i.e. EW $<$ 400\AA). This clustering of AGN at low EW,
as compared to the distribution of EW for the brighter sample, indicates there
is host galaxy light raising the observed continuum luminosity in this
wavelength region for these few objects.

In order to subtract the host galaxy for these few faint objects, we use a
\citet{Bruzual03} model of an old stellar population, with an age of 11 Gyr and
solar metallicity. Such stellar population models have been used in many
earlier works to correct for host galaxy contamination
\citep[e.g.][]{Bongiorno14,Banerji15}. We scale the stellar population model
based on the ratio between the observed H$\alpha$ EW and the median of the EW
distribution (400\AA). Younger stellar populations have a larger contribution
in the UV, but using a stellar population with an age of 900 Myr, instead of
the 11 Gyr model, changes the luminosity by less than 5\% at 3000 \AA\ in the
corrected AGN spectrum. Therefore, the choice of stellar population model does
not have a large effect on the UV spectrum of our AGN. We now use these
corrected spectra for measuring \Mdot\ and for the remainder of the analysis in
this paper.

\subsection{Bayesian SED-Fitting Procedure}
\label{sec:bayes}

\begin{table}
\caption{Parameter values for the grid of AD models.}
\begin{tabular}{ccl}
  \hline
  Parameter & $\Delta$ & Min-Max values \\
  \hline
  $\log M_{BH}$ [\Msun]             & 0.075   & $7.40:10.25$  \\
  $\log \dot{M}$ [\Msunyr]          & 0.075   & $-1.50:+2.10$ \\
  $a_*$                             & 0.1     & $-1.0:+0.998$ \\
  cos$\theta$(1+2cos$\theta$)/3     & 0.067   & $1.000:0.330$ \\
  $A_V$ (mag)                       & 0.05    & $0.00:0.50$ \\
  \hline
\end{tabular}
\label{tab:bay}
\end{table}

We again generate a grid of thin AD models using the \citet{Slone12} code, and
we use a Bayesian method to fit the models to the observed spectra, in order to
take into account the errors in \mbh\ and \Mdot\ and the unknown disc
inclination. We use the same method described in \citetalias{Capellupo15},
except that the grid now extends to lower \mbh\
and we now have a finer spacing in \mbh\ and \Mdot\ values
(0.075 dex, instead of 0.15 dex; see Table \ref{tab:bay}). The expanded grid
now includes 441,441 models.

In \citetalias{Capellupo15}, we explored applying an intrinsic reddening
correction to those AGN spectra that were not initially well fit by the thin AD
model. However, it is
possible that some of the AGN whose spectra are well fit are also affected by
some amount of intrinsic reddening. We therefore add intrinsic reddening as
another parameter in the Bayesian analysis. We adopt a range in $A_V$ from
0. to 0.50 mag, in intervals of 0.05. To minimize the number of parameters, we
adopt only a simple power-law curve, where $A(\lambda)=A_{o}\lambda^{-1}$ mag,
to deredden the X-shooter spectra. To deredden the GALEX photometry, we use the
MRN dust extinction model \citep{Mathis77}.

To summarize our Bayesian approach, we determine the posterior probability for
each of the 441,441 models for each value of $A_V$ for each source. This
probability is the product of the likelihood, $\mathcal{L}(m)$, and the priors
on \mbh\ and \Mdot. We have no prior knowledge on $a_*$,
cos $\theta$\footnote{We only consider cos $\theta$ $>$ 0.5, appropriate for
type-I AGN.},
or the amount of intrinsic reddening.
The likelihood is based on the standard $\chi^2$ statistic, measured using up
to seven line-free continuum windows, centred at 1353, 1464, 2200, 4205, 5100,
6205, and 8600 \AA. The widths of these bands range from 10 to 50 \AA. For five
objects at the upper end of the narrow redshift range of our sample, the bands
centred on 4205 and 5100 \AA\ fall within regions of strong atmospheric
absorption and are thus unusable. When calculating $\chi^2$, we combine the
standard error from Poisson noise and an assumed 5 per cent error on the flux
calibration.

We use Gaussian distributions, centred on the observed values ($M^{obs}_{BH}$,
$\dot{M}^{obs}$) and with standard deviations ($\sigma_M$, $\sigma_{\dot{M}}$)
given by their uncertainties, to represent the priors on $M_{BH}$ and
$\dot{M}$. We again adopt 0.3 and 0.2 dex for $\sigma_M$ and
$\sigma_{\dot{M}}$, respectively. The resulting posterior probability is given
by
\begin{eqnarray*}
  \textrm{posterior} \propto \exp(-\chi^2/2) \times  \exp(-(M^{obs}_{BH}\!-M^{mod}_{BH})^{2}/2\sigma_{M}^{2}) \\
  \times \exp(-(\dot{M}^{obs}\! \times\! \frac{M^{obs}_{BH}}{M^{mod}_{BH}}-\dot{M}^{mod})^{2}/2\sigma_{\dot{M}}^{2}).
\label{eq:bay}
\end{eqnarray*}
Appendix A in \citetalias{Capellupo15} gives the full derivation of the posterior probability.

The Bayesian procedure ranks the 441,441 models based on the posterior
probability for each one. We consider an AGN to have a satisfactory thin AD fit
when the model with the highest probability has a reduced $\chi^2$ statistic
less than 3.

\subsection{Fitting X-shooter Spectra}
\label{sec:xsh_only}

We first fit thin AD models to just the X-shooter spectra for all the sources.
From \citetalias{Capellupo15}, 22 out of 30 AGN have a satisfactory fit, before
making any additional corrections to the spectra (i.e. correcting for intrinsic
reddening or considering disc winds). After correcting for intrinsic reddening,
but using only a single value of $A_V$ per source, we found satisfactory fits to
another 3 out of 30 sources, bringing the total to 25 out of 30 AGN.

Using a larger model grid and considering multiple values of $A_V$, we find
that 37 out of the entire sample of 39 AGN have satisfactory fits. Three of the
AGN with marginal fits in \citetalias{Capellupo15} can now be fit satisfactorily, and all of the
9 AGN we add to the sample in the current work have satisfactory fits. Only one
of the 9 new sources (J1021-0027) requires an intrinsic reddening correction
for a satisfactory fit (in total, six of the 39 sources require such a
correction for a satisfactory fit).
The best-fit models for the 9 new sources are overplotted in
Fig.~\ref{fig:sp1}.

\begin{figure*}
 \centering
 \includegraphics[width=175mm]{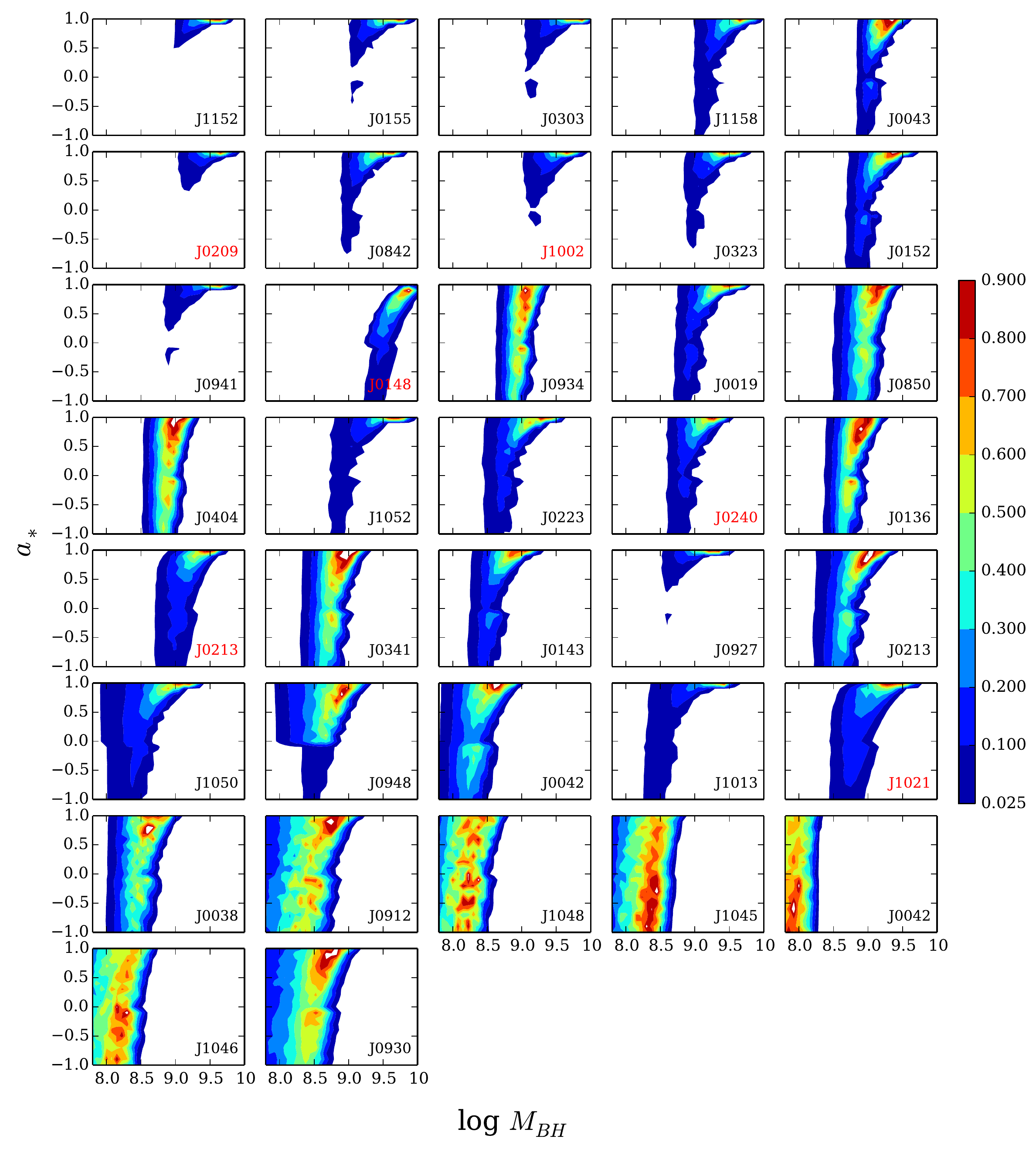}
 \caption{Contour plots of spin parameter $a_*$ versus $M_{BH}$ for the 37
    sources with satisfactory fits to just the X-shooter spectrum.
    The objects labeled with red typeface are those sources which require an
    intrinsic reddening correction to obtain a satisfactory fit.
    The darkest blue contours correspond to a probability of less than 10
    per cent.}
 \label{fig:a_mbh_xsh}
\end{figure*}

In Fig. \ref{fig:a_mbh_xsh}, we show the probability contours for two of the
five parameters, $a_*$ versus
\mbh, for the 37 AGN with satisfactory fits to the X-shooter spectrum. The six
sources that can only be fit after dereddening the spectra are highlighted in
red. Table \ref{tab:master_table} lists the median values of the deduced
parameters based on the probabilities.

\subsection{Fitting X-shooter+GALEX SEDs}
\label{sec:xsh_galex}

While X-shooter provides excellent wavelength coverage, we are missing a
significant portion of the AGN SED that is dominated by emission from the
accretion disc. In particular, we are missing wavelengths blueward of
$\sim$1200 \AA, where, in most cases, a turnover in the thin AD spectrum
occurs. Some constraint on the AGN SED at these short wavelengths is necessary
to fully test the thin AD theory and constrain the various input parameters via
the Bayesian method we adopt.

One solution that is already readily available is the GALEX survey. As
described in Section \ref{sec:galex}, the latest data release of GALEX contains
photometric data for all but one of our sources at $\sim$900\AA, and for 20 out
of 39 at $\sim$600\AA. However, there are two main caveats to the usage of
GALEX photometry. The first is that the GALEX bands are very broad, and we
cannot properly take into account any emission lines or potential intervening
Ly$\alpha$ absorption that could affect the flux at these wavelengths. The
second caveat is variability between the GALEX and X-shooter epochs, especially
given that variability is known to be more significant at these short
wavelengths \citep{MacLeod12,Zuo12}.

With these caveats in mind, we apply our Bayesian method to a combined
X-shooter+GALEX SED. The procedure is the same as in
Section~\ref{sec:xsh_only}, but we now have up to 9 continuum regions, instead
of 7. Because we have multiple epochs of GALEX photometry for most sources, we
use the weighted average of all the epochs for each source. For the error on
each GALEX measurement, we combine the standard measurement errors with an
extra error of 20\% to take into account the unknown variability between the
X-shooter and GALEX epochs and an additional 5\% error based on the unknown
slope of the SED through the GALEX filters. The error estimate for the unknown
variability is based on the typical variability
amplitudes found by \citet{MacLeod12} and \citet{Zuo12} and the variability
between individual GALEX epochs in our own sample.

\begin{figure*}
  \centering
  \includegraphics[width=165mm]{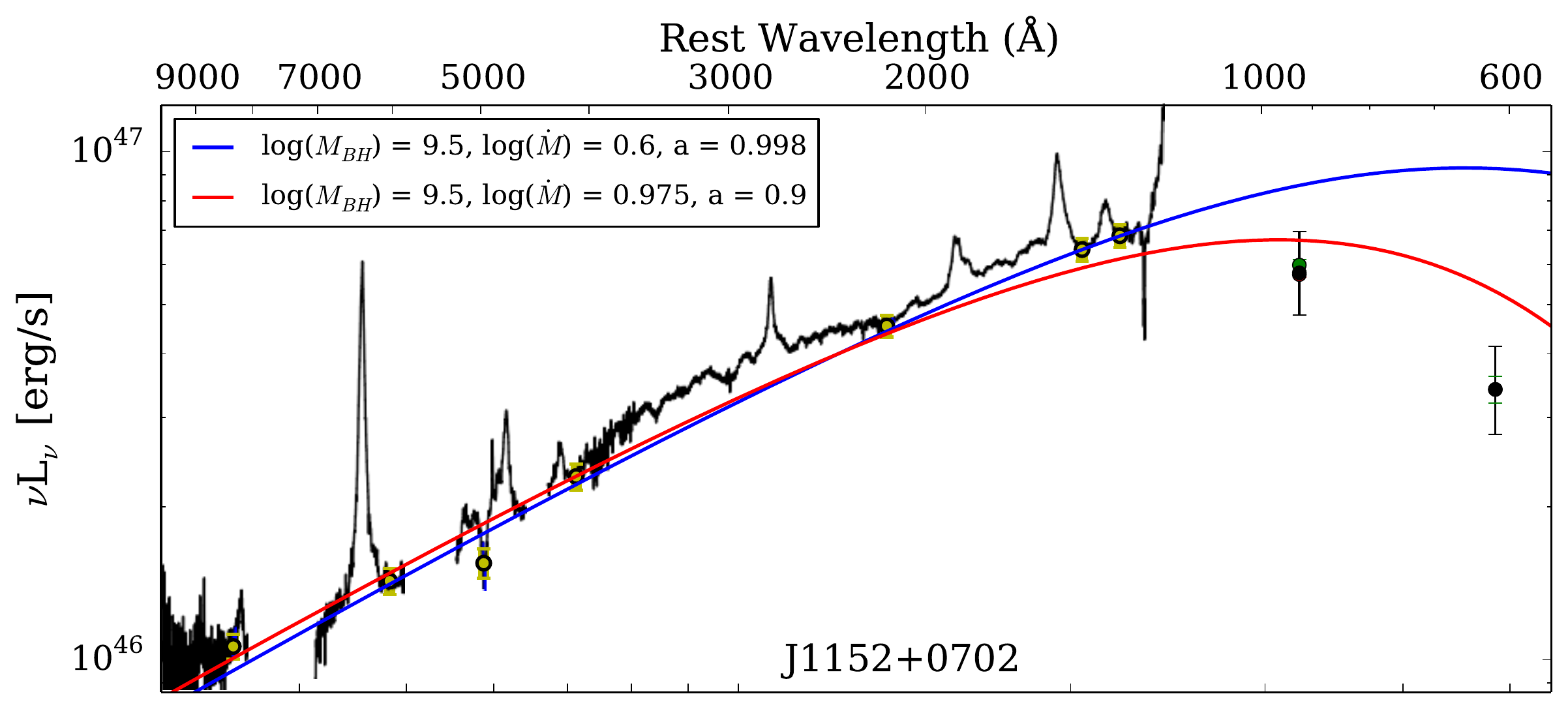}
  \includegraphics[width=165mm]{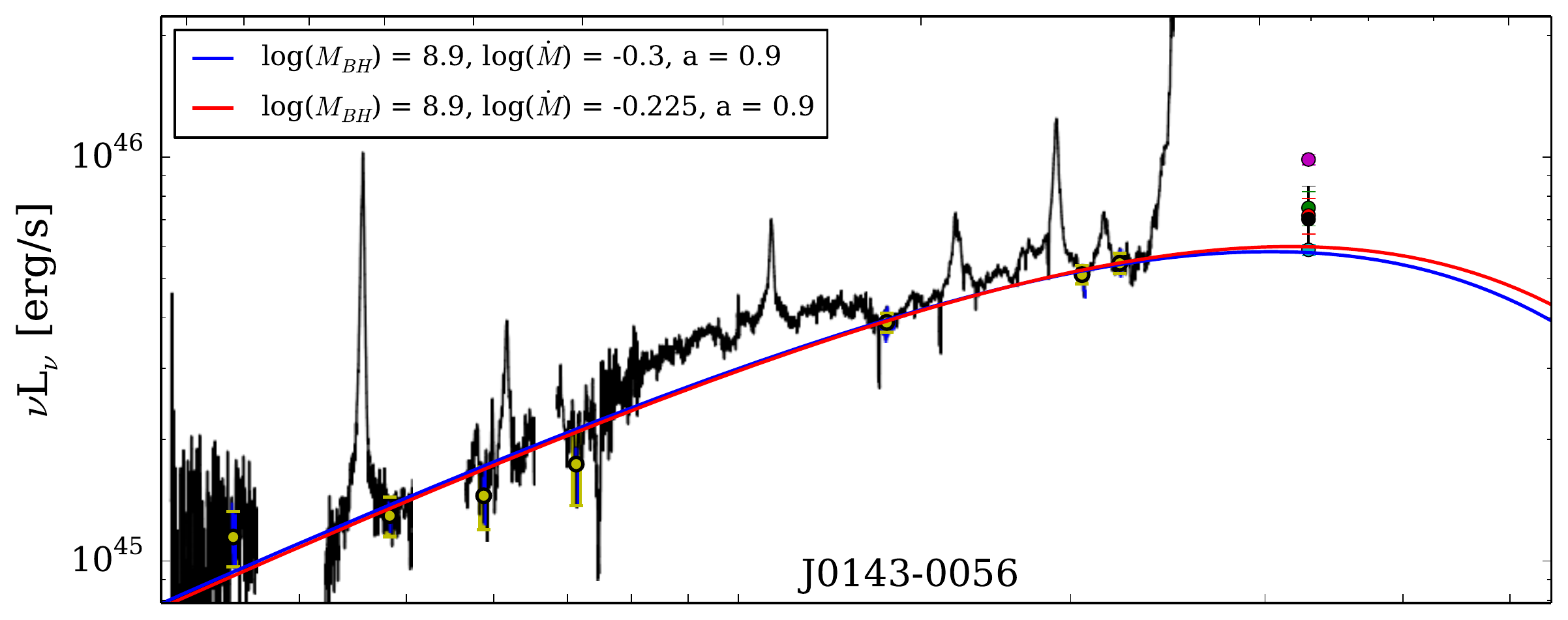}
  \includegraphics[width=165mm]{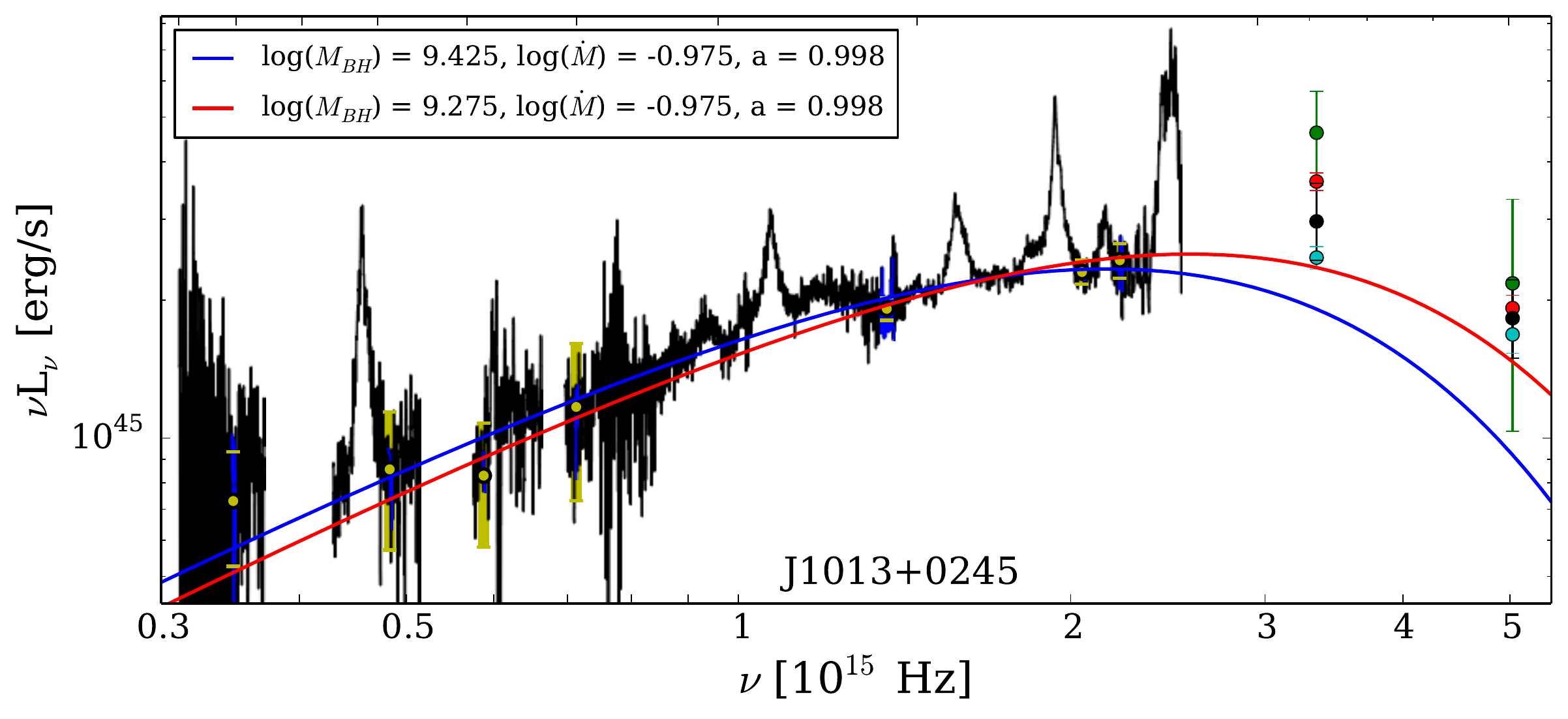}
  \caption{Examples of satisfactory fits to the combined X-shooter+GALEX SED.
    The blue curve is the best-fit to just the X-shooter spectrum, and the red
    curve is the best-fit to X-shooter+GALEX. The colored points are the
    individual GALEX epochs, and the black points are the weighted average of
    the different epochs.}
  \label{fig:gal_sp1}
\end{figure*}

\begin{figure*}
  \centering
  \includegraphics[width=165mm]{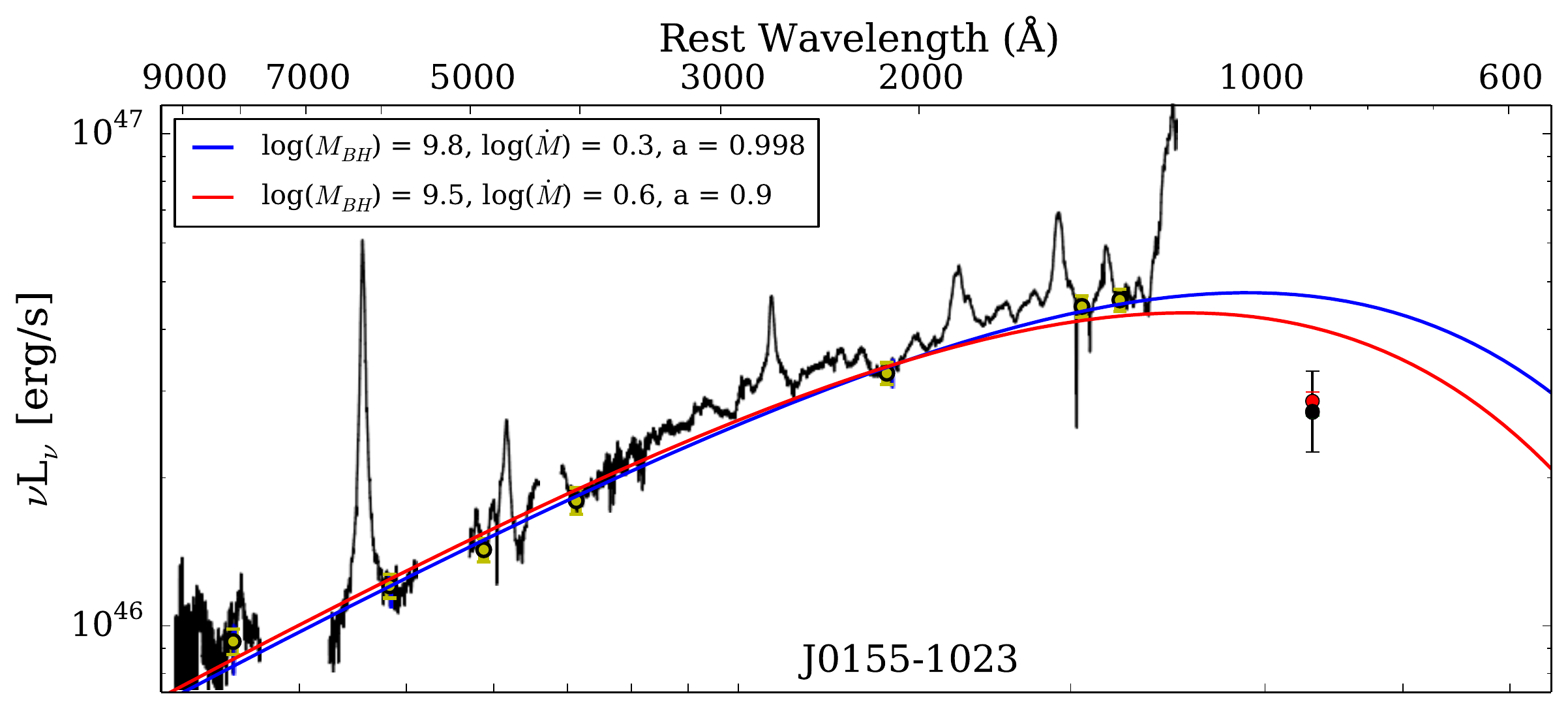}
  \includegraphics[width=165mm]{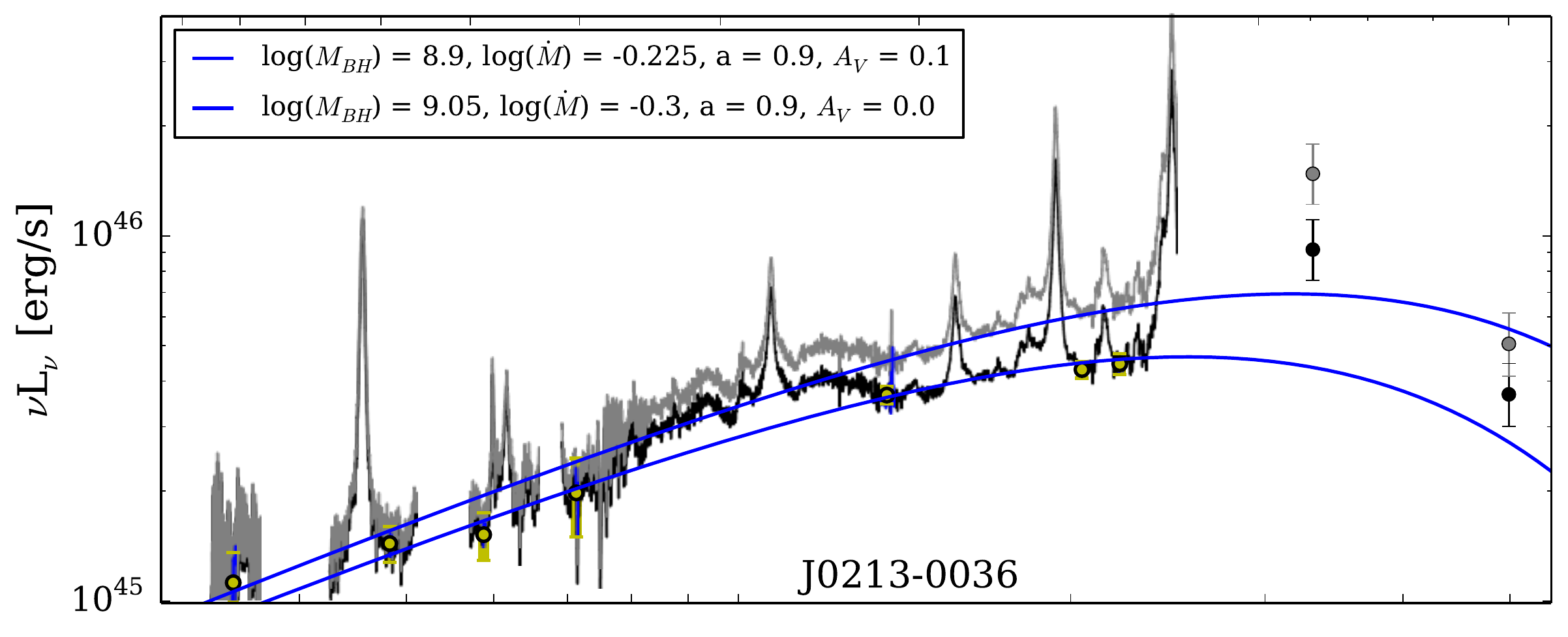}
  \includegraphics[width=165mm]{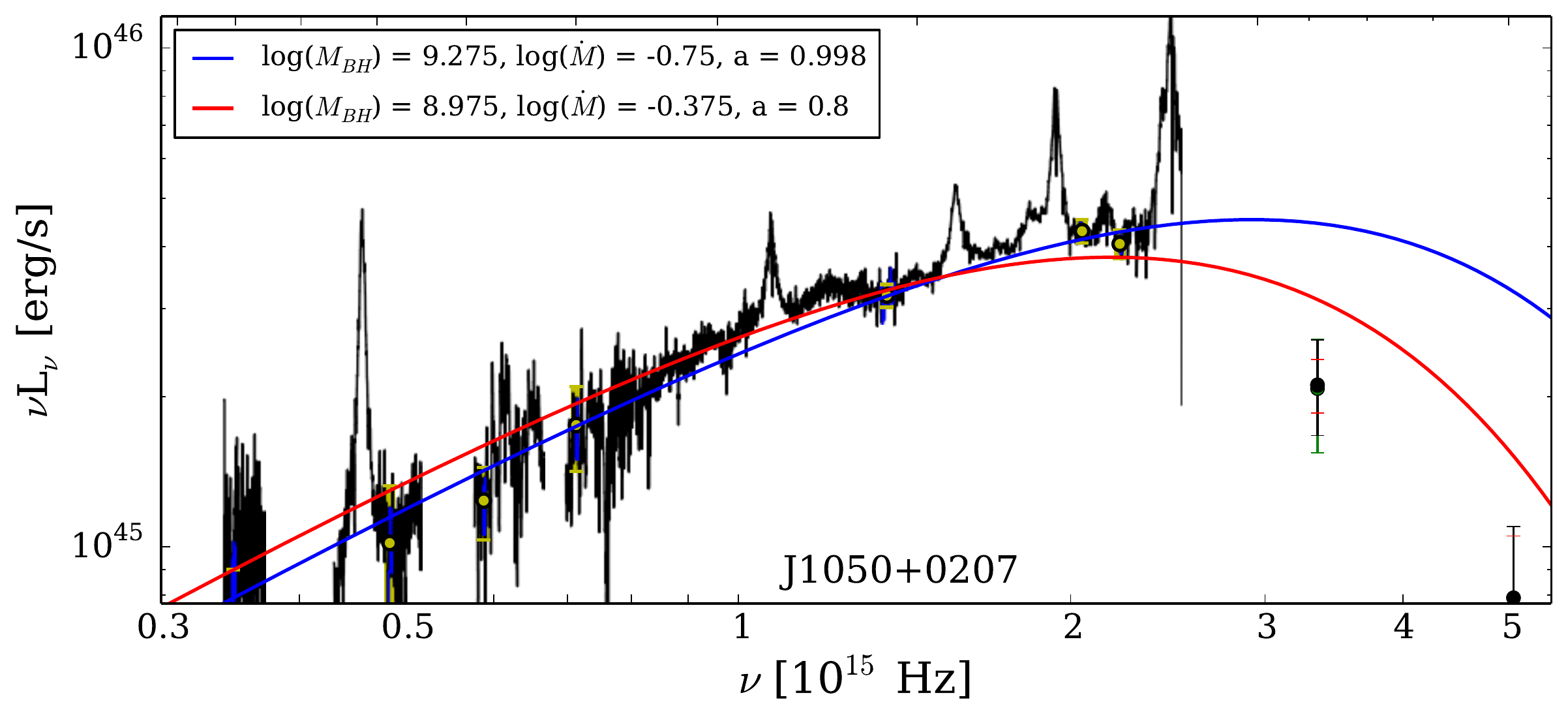}
  \caption{Same as Fig. \ref{fig:gal_sp1}, but for cases where no satisfactory
    fit was found to the X-shooter+GALEX SED. For J0213$-$0036, we show just
    the fits to the X-shooter+GALEX SED, for before and after applying an
    intrinsic reddening correction. The grey curve and points are the
    dereddened SED.}
  \label{fig:gal_sp2}
\end{figure*}
 
In Figs \ref{fig:gal_sp1} and \ref{fig:gal_sp2}, we show several
representative examples of the X-shooter+GALEX SED, with the best-fit model
shown in red and the best-fit model to the X-shooter spectrum alone shown in
blue. The colored points are the individual GALEX epochs, and the black points
are the weighted average of all the epochs. Fig. \ref{fig:gal_sp1} shows three
examples of satisfactory fits, and Fig. \ref{fig:gal_sp2} shows three examples
of cases with a marginal fit or with clearly no fit at all.
We are able to find satisfactory fits to 26/38 of the combined X-shooter+GALEX
SEDs.

Just as in Section~\ref{sec:xsh_only}, we consider intrinsic reddening when
fitting the X-shooter+GALEX SEDs. However, we find that correcting for
intrinsic reddening does not solve the discrepancy we find between the models
and the GALEX photometry for the objects that have satisfactory fits to
X-shooter alone. There are just two sources whose X-shooter+GALEX SEDs are fit
only with $A_V$ > 0, but these are two of the sources that already required
dereddening for a satisfactory fit to the X-shooter spectrum alone.

The examples in Fig. \ref{fig:gal_sp1}, in particular J0143$-$0056 and
J1013+0245, show how variability between the X-shooter and GALEX epochs can
cause the
difference between a good and a bad fit. For example, the magenta GALEX point
for J0143$-$0056 and the green points for J1013+0245 would not be fit with the
thin AD model. If we only had those epochs available, then these two objects
would not be considered to have satisfactory fits. If we had contemporaneous
UV data for J1050+0207 (Fig. \ref{fig:gal_sp2}), for example, it is possible
that we would find a satisfactory fit to the entire SED. Therefore, we can see
from many of the objects with multi-epoch GALEX data that the unknown
variability between the X-shooter and GALEX epochs is a real uncertainty, and
the fraction with satisfactory fits (26/38) is likely a lower limit.

It is also instructive to examine in how many cases our `best-fit' models
overestimate and underestimate the GALEX luminosities. If the discrepancies
between the model and the GALEX measurements are due primarily to variability,
then one would expect to find roughly the same number of cases where the model
overestimates these measurements versus the number where the model
underestimates these measurements. Considering the best-fit model to just the
X-shooter spectrum, roughly the same number overestimate the GALEX photometry
versus underestimate (11 versus 9 sources). Similarly, when fitting the
X-shooter+GALEX SED, half of the best-fit models overestimate the GALEX
luminosities and half underestimate. For this comparison, we are
considering just the weighted average of the GALEX measurements. These results
show that the thin AD model does not systematically overestimate or
underestimate the GALEX data.

\begin{figure*}
  \centering
  \includegraphics[width=175mm]{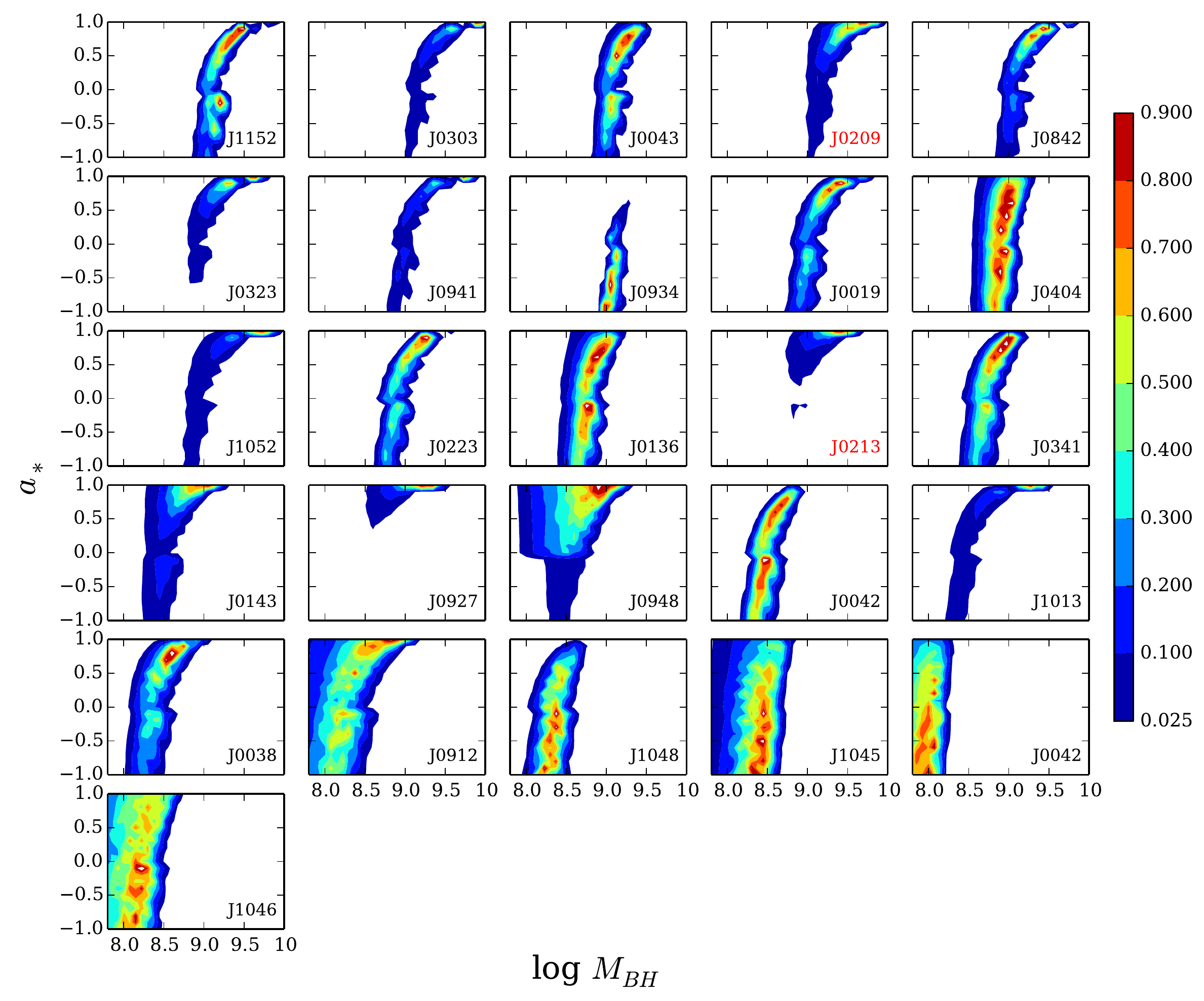}
  \caption{Same as Fig. \ref{fig:a_mbh_xsh}, but for fits to the combined
    X-shooter+GALEX SEDs.}
  \label{fig:a_mbh_gal}
\end{figure*}

As in Section \ref{sec:xsh_only}, we plot the $a_*$ versus \mbh\ probability
contours for fitting the thin AD models now to the X-shooter+GALEX SED. The
constraints on the spin are less confined for many sources when including the
GALEX photometry in the fitting, especially for the AGN with higher \mbh, e.g.
J1152+0702. For some of these high mass cases, the contours are shifted to
lower spin parameters than when fitting the X-shooter spectrum alone.


\begin{landscape}

\begingroup

\begin{table}
\scriptsize
\setlength{\tabcolsep}{3.5pt} 
\caption{Measured and deduced physical parameters}
    \begin{tabular}{cccccccccccccccccccc}
    \hline
    Name & SDSS Coord. Name & log(L3000) & \multicolumn{3}{c}{------ log $M_{BH}$ [M$_{\sun}$] ------} & \multicolumn{3}{c}{------ log $\dot{M}$ [M$_{\sun}$/yr] ------} & log$L/L_{Edd}$ & \multicolumn{2}{c}{----- log $\dot{m}$ -----} & \multicolumn{2}{c}{------ cos$\theta$ ------} & \multicolumn{2}{c}{-------------- $a_*$ --------------} & \multicolumn{2}{c}{----------- $A_{V}$ -----------} & log(L5100) & log($L_{Bol}$)$^{(a)}$ \\
         &                  & [erg/s] & Obs. & Xsh & Xsh-GAL & Obs. & Xsh & Xsh-GAL & [BC] & Xsh & Xsh-GAL & Xsh & Xsh-GAL & Xsh & Xsh-GAL & Xsh & Xsh-GAL & [erg/s] & [erg/s] \\
    \hline
    J1152$+$0702 & J115239.68$+$070222.0 & 46.549 &   9.52 &  9.46$^{+0.20}_{-0.28}$ &  9.18$^{+0.18}_{-0.14}$ &    0.70 &   0.84$^{+0.31}_{-0.21}$ &   1.33$^{+0.30}_{-0.30}$ &   -0.66 &  -0.66 & -0.47 &   0.87 &   0.76 &    0.980$^{+0.048}_{-0.120}$ &    0.152$^{+0.625}_{-0.700}$ &   0.00$^{+0.02}_{-0.02}$ &   0.00$^{+0.02}_{-0.02}$ &   46.082 & 46.866 \\
    J0155$-$1023 & J015504.74$-$102328.4 & 46.428 &   9.54 &  9.45$^{+0.28}_{-0.29}$ &                         &    0.66 &   0.80$^{+0.30}_{-0.32}$ &                          &   -0.81 &  -0.76 &       &   0.87 &        &    0.960$^{+0.061}_{-0.301}$ &                              &   0.05$^{+0.05}_{-0.05}$ &                          &   45.930 & 46.718 \\
    J0303$+$0027 & J030342.79$+$002700.6 & 46.372 &   9.69 &  9.54$^{+0.30}_{-0.36}$ &  9.44$^{+0.46}_{-0.28}$ &    0.44 &   0.71$^{+0.37}_{-0.34}$ &   0.80$^{+0.35}_{-0.55}$ &   -1.01 &  -0.94 & -0.99 &   0.86 &   0.86 &    0.960$^{+0.061}_{-0.434}$ &    0.767$^{+0.224}_{-0.869}$ &   0.08$^{+0.06}_{-0.07}$ &   0.01$^{+0.04}_{-0.03}$ &   46.196 & 46.413 \\
    J1158$-$0322 & J115841.37$-$032239.9 & 46.357 &   9.49 &  9.37$^{+0.29}_{-0.24}$ &                         &    0.58 &   0.83$^{+0.31}_{-0.31}$ &                          &   -0.83 &  -0.84 &       &   0.86 &        &    0.828$^{+0.172}_{-0.776}$ &                              &   0.10$^{+0.07}_{-0.09}$ &                          &   46.127 & 46.646 \\
    J0043$+$0114 & J004315.07$+$011445.8 & 46.272 &   9.10 &  9.15$^{+0.18}_{-0.16}$ &  9.11$^{+0.16}_{-0.12}$ &    0.90 &   0.83$^{+0.25}_{-0.24}$ &   0.94$^{+0.22}_{-0.26}$ &   -0.52 &  -0.66 & -0.75 &   0.87 &   0.86 &    0.772$^{+0.207}_{-0.824}$ &    0.280$^{+0.477}_{-0.748}$ &   0.06$^{+0.05}_{-0.06}$ &   0.01$^{+0.04}_{-0.03}$ &   46.109 & 46.642 \\
    J0209--0947  & J020951.09$-$094727.3 & 46.263 &   9.45 &  9.46$^{+0.22}_{-0.27}$ &  9.39$^{+0.29}_{-0.24}$ &    0.62 &   0.63$^{+0.30}_{-0.25}$ &   0.72$^{+0.30}_{-0.30}$ &   -0.76 &  -0.93 & -0.97 &   0.88 &   0.87 &    0.964$^{+0.059}_{-0.277}$ &    0.827$^{+0.166}_{-0.600}$ &   0.15$^{+0.05}_{-0.09}$ &   0.09$^{+0.08}_{-0.07}$ &   45.792 & 46.566 \\
    J0842$+$0151 & J084240.64$+$015134.1 & 46.226 &   9.39 &  9.29$^{+0.30}_{-0.25}$ &  9.18$^{+0.24}_{-0.17}$ &    0.42 &   0.65$^{+0.30}_{-0.31}$ &   0.78$^{+0.27}_{-0.29}$ &   -0.86 &  -0.86 & -0.88 &   0.86 &   0.84 &    0.900$^{+0.112}_{-0.486}$ &    0.550$^{+0.322}_{-0.870}$ &   0.07$^{+0.05}_{-0.06}$ &   0.01$^{+0.03}_{-0.02}$ &   46.093 & 46.417 \\
    J1002+0331   & J100248.16$+$033155.9 & 46.195 &   9.44 &  9.44$^{+0.24}_{-0.26}$ &                         &    0.53 &   0.64$^{+0.30}_{-0.27}$ &                          &   -0.82 &  -0.98 &       &   0.86 &        &    0.929$^{+0.087}_{-0.466}$ &                              &   0.20$^{+0.06}_{-0.11}$ &                          &   46.112 & 46.643 \\
    J0323--0029  & J032349.53$-$002949.8 & 46.153 &   9.32 &  9.27$^{+0.26}_{-0.26}$ &  9.22$^{+0.37}_{-0.26}$ &    0.44 &   0.62$^{+0.31}_{-0.29}$ &   0.66$^{+0.32}_{-0.41}$ &   -0.73 &  -0.85 & -0.88 &   0.87 &   0.87 &    0.912$^{+0.102}_{-0.490}$ &    0.816$^{+0.171}_{-0.533}$ &   0.10$^{+0.05}_{-0.08}$ &   0.08$^{+0.04}_{-0.05}$ &   45.786 & 46.570 \\
    J0152$-$0839 & J015201.24$-$083958.2 & 46.116 &   9.24 &  9.08$^{+0.28}_{-0.21}$ &                         &    0.44 &   0.71$^{+0.29}_{-0.32}$ &                          &   -0.81 &  -0.75 &       &   0.85 &        &    0.738$^{+0.244}_{-0.944}$ &                              &   0.06$^{+0.06}_{-0.06}$ &                          &   46.033 & 46.421 \\
    J0941$+$0443 & J094126.50$+$044328.8 & 46.083 &   9.59 &  9.36$^{+0.29}_{-0.40}$ &  9.18$^{+0.54}_{-0.23}$ &    0.03 &   0.40$^{+0.39}_{-0.33}$ &   0.59$^{+0.31}_{-0.57}$ &   -1.19 &  -1.06 & -1.03 &   0.85 &   0.85 &    0.966$^{+0.057}_{-0.392}$ &    0.628$^{+0.349}_{-0.992}$ &   0.05$^{+0.06}_{-0.05}$ &   0.00$^{+0.02}_{-0.02}$ &   45.866 & 46.188 \\
    J0148+0003   & J014812.83$+$000322.9 & 46.059 &   9.61 &  9.65$^{+0.22}_{-0.21}$ &                         &    0.36 &   0.69$^{+0.29}_{-0.27}$ &                          &   -0.91 &  -1.42 &       &   0.85 &        &    0.585$^{+0.299}_{-0.669}$ &                              &   0.44$^{+0.06}_{-0.11}$ &                          &   45.676 & 46.481 \\
    J0934$+$0005 & J093411.15$+$000519.7 & 45.939 &   8.84 &  8.97$^{+0.13}_{-0.14}$ &  9.07$^{+0.08}_{-0.06}$ &    0.66 &   0.64$^{+0.23}_{-0.21}$ &   0.59$^{+0.14}_{-0.11}$ &   -0.58 &  -0.91 & -1.25 &   0.88 &   0.91 &    0.280$^{+0.558}_{-0.800}$ &   -0.551$^{+0.460}_{-0.348}$ &   0.10$^{+0.09}_{-0.08}$ &   0.00$^{+0.02}_{-0.02}$ &   45.829 & 46.150 \\
    J0019$-$1053 & J001946.98$-$105313.4 & 45.791 &   9.33 &  9.14$^{+0.33}_{-0.25}$ &  9.09$^{+0.26}_{-0.16}$ &    0.03 &   0.27$^{+0.32}_{-0.35}$ &   0.35$^{+0.25}_{-0.33}$ &   -1.21 &  -1.20 & -1.30 &   0.85 &   0.85 &    0.797$^{+0.200}_{-0.948}$ &    0.346$^{+0.513}_{-0.869}$ &   0.09$^{+0.10}_{-0.08}$ &   0.02$^{+0.05}_{-0.03}$ &   45.991 & 46.048 \\
    J0850$+$0022 & J085027.88$+$002255.0 & 45.742 &   8.89 &  8.96$^{+0.17}_{-0.19}$ &                         &    0.51 &   0.52$^{+0.28}_{-0.26}$ &                          &   -0.82 &  -1.01 &       &   0.86 &        &    0.293$^{+0.580}_{-0.836}$ &                              &   0.20$^{+0.11}_{-0.12}$ &                          &   45.945 & 46.271 \\
    J0404$-$0446 & J040414.14$-$044649.8 & 45.729 &   8.76 &  8.90$^{+0.14}_{-0.15}$ &  8.90$^{+0.13}_{-0.13}$ &    0.65 &   0.56$^{+0.24}_{-0.21}$ &   0.56$^{+0.23}_{-0.21}$ &   -0.70 &  -0.92 & -0.97 &   0.87 &   0.87 &    0.291$^{+0.571}_{-0.824}$ &    0.099$^{+0.619}_{-0.711}$ &   0.15$^{+0.09}_{-0.09}$ &   0.13$^{+0.09}_{-0.08}$ &   45.177 & 46.493 \\
    J1052$+$0236 & J105213.24$+$023606.0 & 45.722 &   9.59 &  9.26$^{+0.42}_{-0.38}$ &  9.28$^{+0.44}_{-0.37}$ &   -0.38 &   0.10$^{+0.41}_{-0.45}$ &   0.07$^{+0.42}_{-0.46}$ &   -1.54 &  -1.39 & -1.50 &   0.85 &   0.85 &    0.897$^{+0.117}_{-0.933}$ &    0.854$^{+0.156}_{-0.975}$ &   0.11$^{+0.11}_{-0.09}$ &   0.09$^{+0.08}_{-0.07}$ &   45.119 & 45.746 \\
    J0223$-$0007 & J022321.39$-$000733.7 & 45.681 &   9.17 &  8.90$^{+0.35}_{-0.25}$ &  8.95$^{+0.23}_{-0.15}$ &   -0.24 &   0.28$^{+0.32}_{-0.36}$ &   0.31$^{+0.23}_{-0.28}$ &   -1.16 &  -0.99 & -1.23 &   0.85 &   0.84 &    0.749$^{+0.239}_{-0.886}$ &    0.254$^{+0.536}_{-0.804}$ &   0.06$^{+0.08}_{-0.06}$ &   0.00$^{+0.02}_{-0.02}$ &   45.084 & 45.970 \\
    J0240--0758  & J024028.85$-$075843.5 & 45.678 &   9.04 &  8.98$^{+0.25}_{-0.21}$ &                         &    0.12 &   0.24$^{+0.29}_{-0.29}$ &                          &   -0.94 &  -1.10 &       &   0.86 &        &    0.756$^{+0.233}_{-0.888}$ &                              &   0.12$^{+0.09}_{-0.09}$ &                          &   45.404 & 46.196 \\
    J0136$-$0015 & J013652.44$-$001524.5 & 45.650 &   8.75 &  8.76$^{+0.17}_{-0.16}$ &  8.77$^{+0.15}_{-0.13}$ &    0.25 &   0.36$^{+0.25}_{-0.26}$ &   0.37$^{+0.23}_{-0.24}$ &   -0.77 &  -0.94 & -1.03 &   0.86 &   0.86 &    0.379$^{+0.494}_{-0.858}$ &    0.082$^{+0.636}_{-0.685}$ &   0.06$^{+0.08}_{-0.06}$ &   0.04$^{+0.07}_{-0.04}$ &   45.437 & 45.717 \\
    J0213--1003  & J021350.45$-$100300.4 & 45.616 &   9.11 &  9.16$^{+0.26}_{-0.23}$ &  9.22$^{+0.24}_{-0.31}$ &    0.29 &   0.33$^{+0.32}_{-0.30}$ &   0.28$^{+0.34}_{-0.28}$ &   -0.84 &  -1.25 & -1.07 &   0.86 &   0.86 &    0.658$^{+0.318}_{-0.921}$ &    0.956$^{+0.064}_{-0.401}$ &   0.38$^{+0.10}_{-0.15}$ &   0.46$^{+0.04}_{-0.08}$ &   45.281 & 46.123 \\
    J0341$-$0037 & J034156.07$-$003706.4 & 45.572 &   8.76 &  8.77$^{+0.18}_{-0.18}$ &  8.75$^{+0.18}_{-0.15}$ &    0.25 &   0.32$^{+0.27}_{-0.27}$ &   0.34$^{+0.25}_{-0.27}$ &   -0.84 &  -0.99 & -1.01 &   0.86 &   0.85 &    0.384$^{+0.511}_{-0.876}$ &    0.213$^{+0.529}_{-0.760}$ &   0.11$^{+0.09}_{-0.08}$ &   0.09$^{+0.06}_{-0.06}$ &   45.184 & 45.804 \\
    J0143$-$0056 & J014334.89$-$005635.3 & 45.534 &   8.83 &  8.65$^{+0.27}_{-0.21}$ &  8.66$^{+0.28}_{-0.23}$ &   -0.03 &   0.18$^{+0.29}_{-0.34}$ &   0.16$^{+0.31}_{-0.33}$ &   -0.95 &  -0.88 & -0.85 &   0.85 &   0.85 &    0.687$^{+0.289}_{-1.005}$ &    0.768$^{+0.223}_{-0.988}$ &   0.04$^{+0.06}_{-0.04}$ &   0.05$^{+0.06}_{-0.05}$ &   45.151 & 45.817 \\
    J0927$+$0004 & J092715.49$+$000401.1 & 45.514 &   9.25 &  9.06$^{+0.27}_{-0.37}$ &  9.06$^{+0.26}_{-0.37}$ &   -0.36 &  -0.15$^{+0.37}_{-0.31}$ &  -0.15$^{+0.36}_{-0.30}$ &   -1.38 &  -1.28 & -1.27 &   0.85 &   0.85 &    0.972$^{+0.053}_{-0.278}$ &    0.976$^{+0.050}_{-0.221}$ &   0.05$^{+0.06}_{-0.05}$ &   0.05$^{+0.06}_{-0.05}$ &   45.430 & 45.730 \\
    J0213$-$0036 & J021310.33$-$003620.5 & 45.487 &   8.92 &  8.74$^{+0.26}_{-0.22}$ &                         &   -0.15 &   0.19$^{+0.30}_{-0.33}$ &                          &   -1.09 &  -1.06 &       &   0.85 &        &    0.468$^{+0.454}_{-0.937}$ &                              &   0.07$^{+0.09}_{-0.07}$ &                          &   45.322 & 45.738 \\
    J1050$+$0207 & J105057.09$+$020708.5 & 45.402 &   9.01 &  8.57$^{+0.41}_{-0.32}$ &                         &   -0.60 &   0.14$^{+0.38}_{-0.42}$ &                          &   -1.21 &  -0.84 &       &   0.84 &        &    0.673$^{+0.308}_{-0.873}$ &                              &   0.06$^{+0.07}_{-0.06}$ &                          &   45.640 & 45.820 \\
    J0948$+$0137 & J094801.42$+$013716.2 & 45.317 &   8.83 &  8.69$^{+0.24}_{-0.29}$ &  8.63$^{+0.28}_{-0.31}$ &   -0.64 &   0.00$^{+0.35}_{-0.29}$ &   0.06$^{+0.36}_{-0.33}$ &   -1.15 &  -1.12 & -0.96 &   0.85 &   0.85 &    0.633$^{+0.295}_{-0.467}$ &    0.704$^{+0.261}_{-0.495}$ &   0.14$^{+0.12}_{-0.11}$ &   0.17$^{+0.09}_{-0.12}$ &   45.295 & 45.398 \\
    J0042$+$0008 & J004213.01$+$000807.3 & 45.260 &   8.52 &  8.36$^{+0.25}_{-0.23}$ &  8.48$^{+0.16}_{-0.12}$ &   -0.14 &   0.04$^{+0.30}_{-0.33}$ &  -0.02$^{+0.21}_{-0.25}$ &   -0.91 &  -0.83 & -1.18 &   0.85 &   0.85 &    0.451$^{+0.481}_{-0.944}$ &   -0.090$^{+0.702}_{-0.601}$ &   0.04$^{+0.07}_{-0.05}$ &   0.01$^{+0.04}_{-0.02}$ &   45.029 & 45.429 \\
    J1013$+$0245 & J101325.50$+$024521.5 & 45.210 &   9.47 &  8.82$^{+0.52}_{-0.38}$ &  8.81$^{+0.49}_{-0.38}$ &   -1.10 &  -0.29$^{+0.42}_{-0.49}$ &  -0.30$^{+0.45}_{-0.46}$ &   -1.89 &  -1.38 & -1.40 &   0.84 &   0.84 &    0.864$^{+0.147}_{-1.054}$ &    0.845$^{+0.165}_{-1.073}$ &   0.07$^{+0.10}_{-0.07}$ &   0.08$^{+0.07}_{-0.06}$ &   45.521 & 45.300 \\
    J1021$-$0027 & J102122.23$-$002723.6 & 45.059 &   9.19 &  9.15$^{+0.32}_{-0.37}$ &                         &   -0.57 &  -0.45$^{+0.41}_{-0.34}$ &                          &   -1.77 &  -1.84 &       &   0.85 &        &    0.880$^{+0.131}_{-0.819}$ &                              &   0.39$^{+0.09}_{-0.15}$ &                          &   45.618 & 45.379 \\
    J0038$-$0019 & J003854.41$-$001926.0 & 45.021 &   8.49 &  8.51$^{+0.19}_{-0.19}$ &  8.45$^{+0.21}_{-0.17}$ &   -0.21 &  -0.28$^{+0.28}_{-0.27}$ &  -0.23$^{+0.29}_{-0.29}$ &   -1.08 &  -1.31 & -1.19 &   0.86 &   0.85 &    0.433$^{+0.488}_{-0.904}$ &    0.461$^{+0.394}_{-0.873}$ &   0.14$^{+0.11}_{-0.10}$ &   0.15$^{+0.08}_{-0.08}$ &   44.782 & 45.589 \\
    J0912$-$0040 & J091216.68$-$004046.4 & 45.011 &   8.64 &  8.47$^{+0.26}_{-0.30}$ &  8.32$^{+0.39}_{-0.31}$ &   -0.34 &  -0.28$^{+0.36}_{-0.33}$ &  -0.18$^{+0.42}_{-0.44}$ &   -1.25 &  -1.34 & -1.01 &   0.85 &   0.85 &    0.249$^{+0.634}_{-0.844}$ &    0.484$^{+0.477}_{-1.000}$ &   0.09$^{+0.12}_{-0.08}$ &   0.15$^{+0.11}_{-0.10}$ &   44.871 & 44.955 \\
    J1048$-$0019 & J104807.57$-$001943.3 & 44.947 &   8.24 &  8.21$^{+0.19}_{-0.23}$ &  8.34$^{+0.12}_{-0.12}$ &   -0.08 &  -0.25$^{+0.30}_{-0.27}$ &  -0.35$^{+0.21}_{-0.20}$ &   -0.90 &  -1.10 & -1.40 &   0.86 &   0.87 &    0.102$^{+0.708}_{-0.737}$ &   -0.234$^{+0.643}_{-0.541}$ &   0.08$^{+0.09}_{-0.07}$ &   0.03$^{+0.07}_{-0.04}$ &   44.713 & 44.967 \\
    J1045$-$0047 & J104549.63$-$004755.3 & 44.862 &   8.23 &  8.34$^{+0.16}_{-0.21}$ &  8.40$^{+0.13}_{-0.16}$ &   -0.29 &  -0.29$^{+0.31}_{-0.24}$ &  -0.33$^{+0.29}_{-0.22}$ &   -0.95 &  -1.31 & -1.43 &   0.87 &   0.88 &   -0.070$^{+0.774}_{-0.641}$ &   -0.206$^{+0.700}_{-0.571}$ &   0.16$^{+0.12}_{-0.11}$ &   0.15$^{+0.13}_{-0.11}$ &   44.361 & 44.852 \\
    J0042$-$0011 & J004230.44$-$001102.3 & 44.806 &   7.69 &  7.91$^{+0.15}_{-0.18}$ &  7.96$^{+0.12}_{-0.14}$ &    0.37 &  -0.07$^{+0.30}_{-0.25}$ &  -0.08$^{+0.29}_{-0.23}$ &   -0.43 &  -0.66 & -0.76 &   0.88 &   0.89 &   -0.147$^{+0.800}_{-0.613}$ &   -0.282$^{+0.694}_{-0.533}$ &   0.22$^{+0.09}_{-0.09}$ &   0.23$^{+0.10}_{-0.10}$ &   44.785 & 45.114 \\
    J1046$+$0025 & J104629.66$+$002538.9 & 44.757 &   8.17 &  8.17$^{+0.18}_{-0.23}$ &  8.16$^{+0.18}_{-0.22}$ &   -0.16 &  -0.40$^{+0.31}_{-0.26}$ &  -0.39$^{+0.31}_{-0.26}$ &   -0.95 &  -1.22 & -1.20 &   0.86 &   0.86 &    0.008$^{+0.764}_{-0.675}$ &    0.023$^{+0.747}_{-0.681}$ &   0.09$^{+0.10}_{-0.08}$ &   0.09$^{+0.10}_{-0.08}$ &   44.310 & 44.848 \\
    J0930$-$0018 & J093046.79$-$001825.8 & 44.701 &   8.55 &  8.47$^{+0.23}_{-0.28}$ &                         &   -0.49 &  -0.56$^{+0.34}_{-0.30}$ &                          &   -1.36 &  -1.63 &       &   0.85 &        &    0.228$^{+0.624}_{-0.818}$ &                              &   0.14$^{+0.15}_{-0.11}$ &                          &   44.416 & 44.704 \\
    \hline
    J1108+0141   & J115841.37$-$032239.9 & 46.337 & 9.37 & & & 0.74 & & & -0.59 & & & & & & & & & 44.625 & \\
    J1005+0245   & J100513.75$+$024510.3 & 46.062 & 9.45 & & & 0.52 & & & -1.07 & & & & & & & & & 44.570 & \\
    \hline
    \end{tabular}
    $(a)$ {Based on the best-fit model to the X-shooter spectrum.}
  \label{tab:master_table}
\end{table}

\endgroup

\end{landscape}


\section{Discussion}
\label{sec:discuss}

\subsection{AGN Accretion Discs}

In this work, we fit standard thin AD models (Section \ref{sec:thinad}) to
X-shooter spectra of 39 AGN at $z$$\sim$1.5 and also to the combined
X-shooter+GALEX SED of 38 of these sources.
When considering just the X-shooter spectrum, we can fit 37 out of 39 AGN
spectra in our sample when allowing for a small intrinsic reddening correction.
\citet{Collinson15} also find agreement between the thin AD model and the
optical/IR spectra for many of their 11 sources.

When including GALEX photometry in our fitting procedure, the number of AGN
that we can fit satisfactorily is reduced to 26 out of 38 AGN.
Accurately fitting SEDs to X-shooter and GALEX data is hampered by
potential variability between the X-shooter and GALEX epochs.
For the sample overall, in roughly half the cases where we do not find a model
fit that is consistent with both the X-shooter spectrum and the GALEX
photometry, the model fit overestimates the GALEX measurements
(See Section \ref{sec:xsh_galex}). Therefore, there is an even
split between overestimating and underestimating the GALEX measurements, indicating
that variability is a likely cause for the discrepancy between model and observations
for the 11 sources that no longer have a satisfactory thin AD model fit.

However, if we consider just the AGN with \mbh\ $>10^{9}$ \Msun, and ignore the
two AGN with broad absorption, the tendency is for the model to overestimate
the GALEX photometry for those cases with no satisfactory fit.
This at least suggests that the discrepancy between the thin AD model and the
GALEX photometry might not be due solely to variability between the GALEX and
X-shooter epochs, at least for the brighter half of the sample, but rather that
there is some physical explanation for the discrepancy.

While we found both in \citetalias{Capellupo15} and in the current work that an
intrinsic reddening correction can cure discrepancies between the model and the
X-shooter spectrum in the bluer part of the X-shooter spectrum, we do not find
that intrinsic reddening helps to cure the discrepancies between the model and
the GALEX photometry mentioned above when our models overestimate the GALEX
luminosities.

One possibility for the discrepancy at short wavelengths is outflowing gas from
the accretion disc. Both \citet{Slone12} and \citet{Laor14} show how including
a mass outflow from a thin AD reduces the radiation at shorter wavelengths,
and this could explain the discrepancy between the data and the model for
those cases where the model overestimates the GALEX photometry.

Another further possibility is that some of these systems do not harbour a
thin AD, but rather a `slim' accretion disc. Such discs are expected at larger
\Ledd\ (\Ledd\ $>\sim0.2$; \citealt{Abramowicz88,Ohsuga11,Netzer13,Wang14}).
However, current models of `slim' discs are not yet able to produce predicted
SEDs that are accurate enough for a comparison to observed SEDs as we perform
in this work \citep[see e.g.][]{Sadowski15}. It will be informative to compare
such model SEDs, when they are available, to datasets like the one presented
here to test what fraction of AGN are consistent with having a `slim' AD.

\subsection{Reddening in AGN Host Galaxies}

\begin{figure}
 \centering
 \includegraphics[width=80mm]{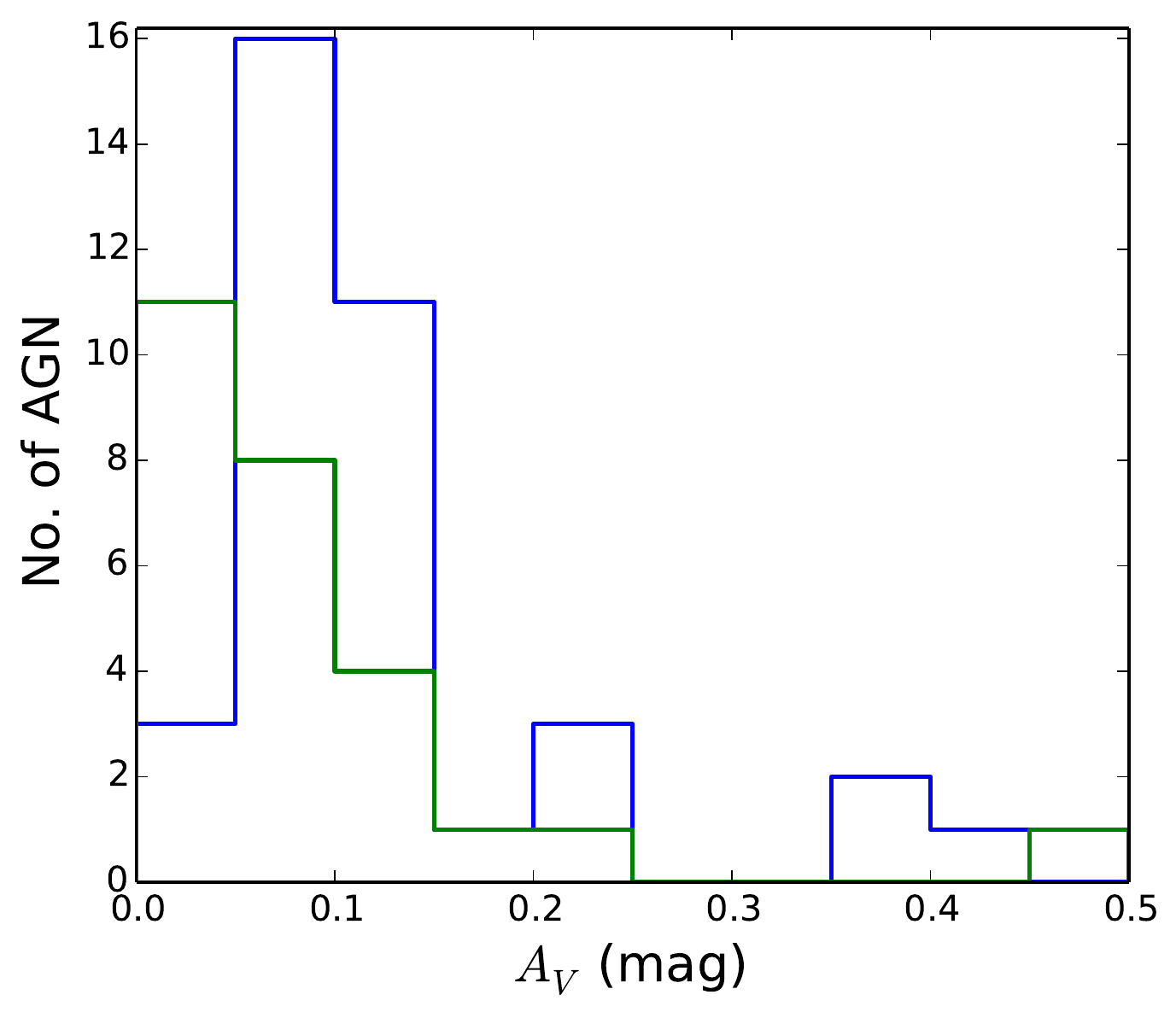}
 \caption{The distribution in the median $A_V$ values from the Bayesian fitting
   procedure. The blue curve is based on fits to the X-shooter spectra alone,
   and the green curve is based on fits to the combined X-shooter+GALEX SED.}
 \label{fig:Av_hist}
\end{figure}

In \citetalias{Capellupo15}, we compared three different extinction curves $-$
simple power-law, Galactic, and SMC $-$ and found that the simple power-law and
Galactic curves gave the best fits to the observed SEDs. In this paper, to
reduce the number of free parameters in our Bayesian fitting procedure, we only
consider the simple power-law model, but we can compare our results to the
typical amount of reddening found in AGN in other work.

In Figure \ref{fig:Av_hist}, we plot the distribution in $A_V$ values from our
Bayesian fitting routine for all the AGN with satisfactory thin AD fits. Most
of the AGN have $A_V$ values $\le 0.15$ mag. For comparison, \citet{Krawczyk15}
find that just 2.5\% of non-BAL quasars, out of a large sample of SDSS quasars,
have $A_V > 0.3$ mag. In our smaller sample, the results of our Bayesian
fitting routine gives 2 out of 37 non-BAL AGN (5\%) with $A_V > 0.3$ mag. This
is generally consistent with the results of \citet{Krawczyk15} and indicates
that, in general, we are not overcorrecting the spectra when including
intrinsic reddening as a parameter in the fitting routine.

While our sample was selected to avoid AGN with significant absorption, there
are two sources in the sample with BAL absorption (J1005+0245 and
J1021$-$0027). We could not find satisfactory fits for either of these two
sources before intrinsic reddening correction, even when fitting the X-shooter
spectrum alone. After correcting for intrinsic reddening, we find a
satisfactory fit for one and a marginal fit for the other. This is consistent
with previous work that has shown that BAL quasars tend to have redder spectra
than non-BAL quasars. For example, \citet{Krawczyk15} find that 13\% of BAL
quasars have $A_V > 0.3$ mag, compared to just 2.5\% of non-BAL quasars, as
mentioned above. One of the BAL AGN in our sample, J1021$-$0027, has
$A_V = 0.39^{+0.09}_{-0.15}$ mag. The other, J1005+0245, does not have a
satisfactory thin AD fit, but the closest fit we find is with an
$A_V$ = 0.50 mag.

\subsection{Disc-Derived \mbh\ and \lledd\ and Bolometric Correction Factors}

\begin{figure}
 \centering
 \includegraphics[width=80mm]{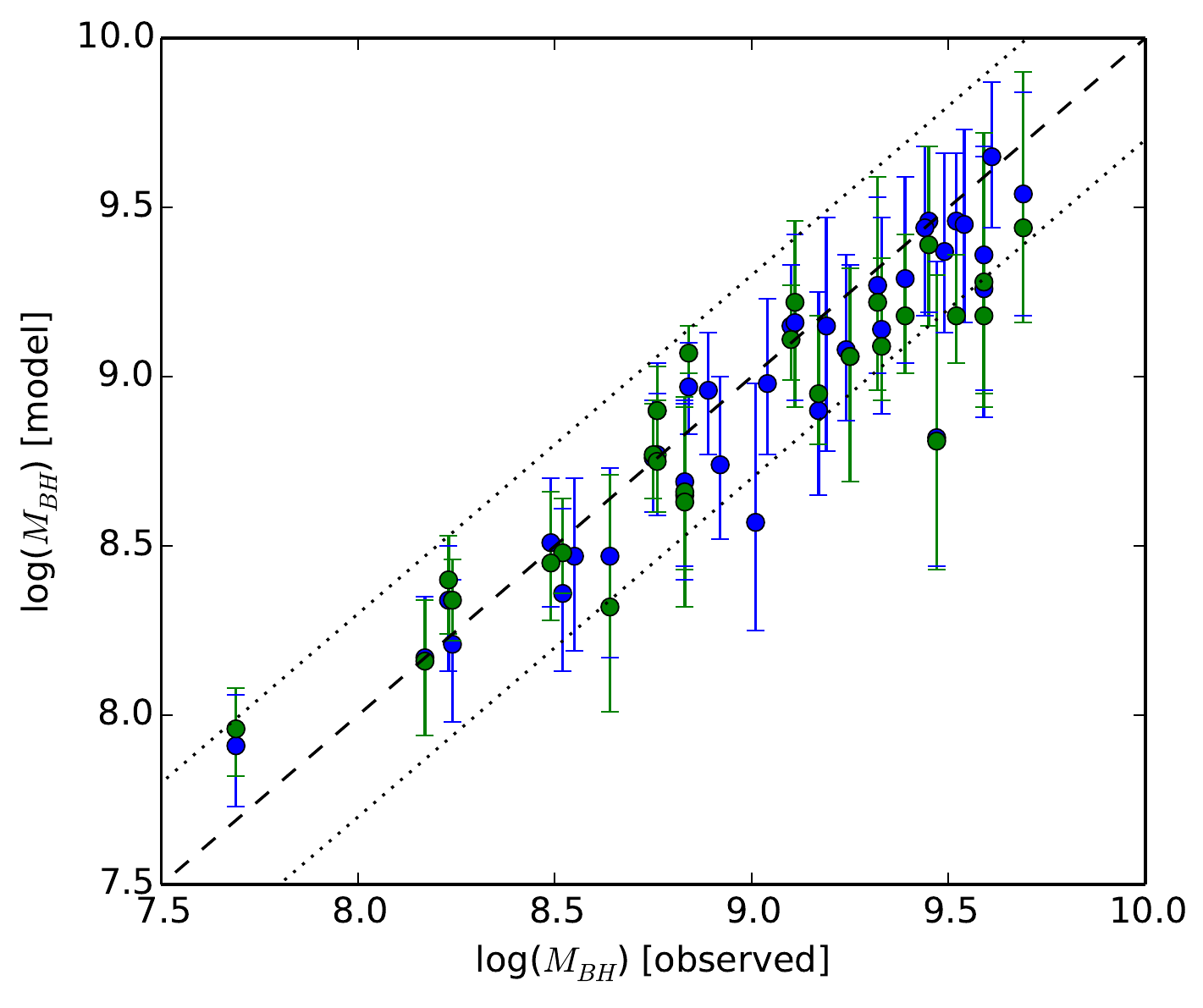}
 \caption{A comparison between the observed \mbh, measured in Paper II directly
   from the spectra, and the median value of \mbh\ from the Bayesian fitting
   procedure for just the X-shooter spectra (blue points) and for the combined
   X-shooter+GALEX SED (green points). For reference, the dashed line is the
   one-to-one line, and the dotted lines are $\pm$0.3 dex. The typical error on
   log($M_{BH}^{obs}$) is 0.3 dex.}
 \label{fig:mbh_comp}
\end{figure}

\begin{figure}
 \centering
 \includegraphics[width=80mm]{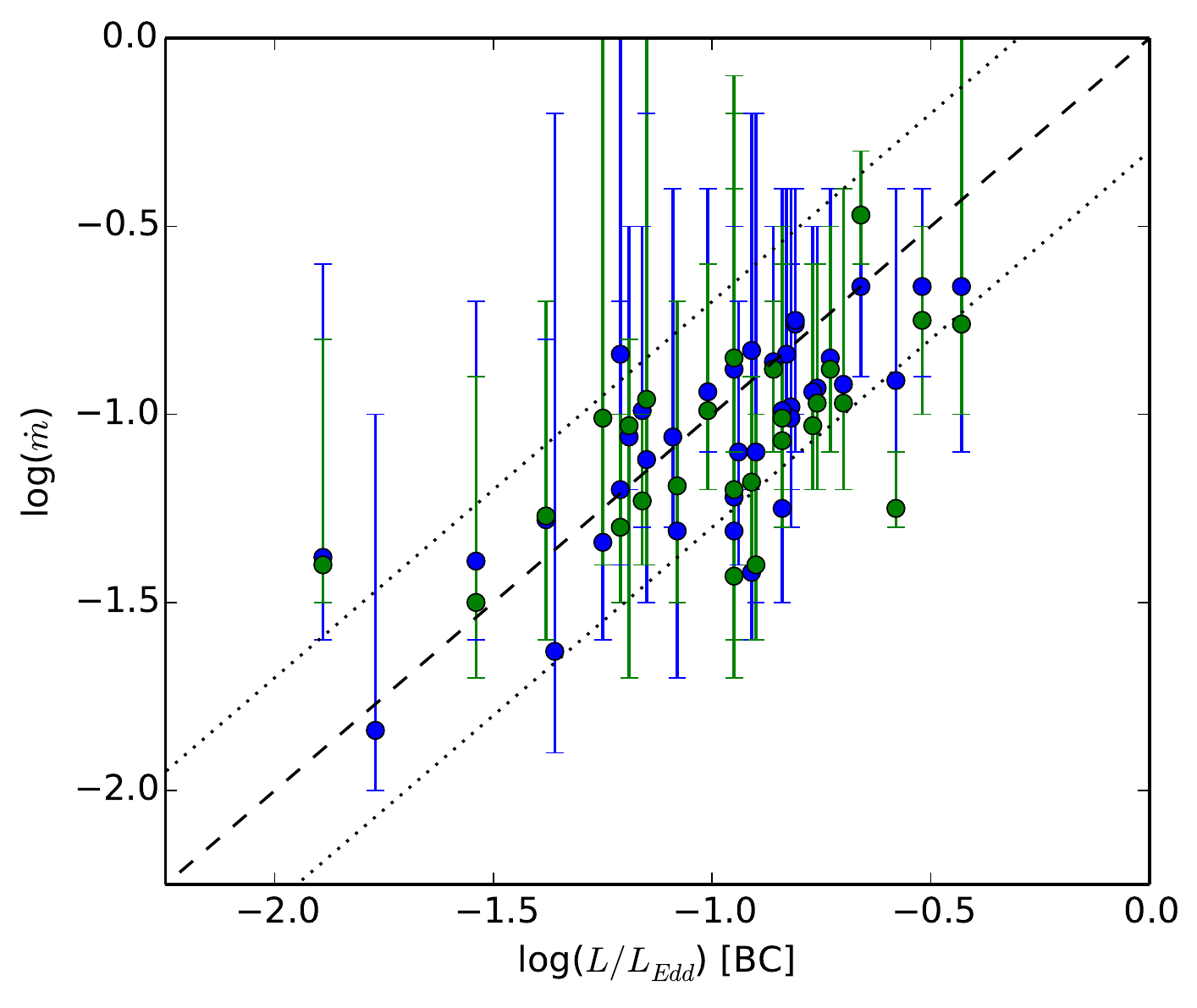}
 \caption{Same as Fig. \ref{fig:mbh_comp}, but instead showing a comparison
   between \lledd[BC], calculated directly from the observed spectra using a
   bolometric correction (BC) factor (Paper II), and the median \mdot\ value
   from the Bayesian fitting routine. The typical errors on \lledd\ are at
   least as high as those on $M_{BH}^{obs}$.}
 \label{fig:mdot_comp}
\end{figure}

Given the fitted thin AD SEDs, we can now compare the values of \mbh\ and
\lledd\ derived from the thin AD fits to our best estimates of \mbh\ and
\lledd\ derived directly from the observed spectrum (Paper II).
In particular, Fig.~\ref{fig:mbh_comp} shows that we are able to find
satisfactory fits for most of the AGN in our sample with thin AD models that
have BH masses within $\sim$1$\sigma$ of the observed values of \mbh.
Interestingly, we also find good agreement between \lledd[BC], which is
measured directly from the observed spectrum using a bolometric correction (BC)
factor, and the median value of \lledd\ (\mdot) from our thin AD fitting
procedure, as shown in Fig.~\ref{fig:mdot_comp}. Comparing
Figs~\ref{fig:mbh_comp} and \ref{fig:mdot_comp} to the corresponding figures in
Paper I, it is clear that we find better agreement here between the results of
the Bayesian analysis and the observationally-derived quantities, especially
between \mdot\ and \lledd[BC]. This is largely due to the improvements in the
measurements of \mbh, as described in Paper II. In Paper I, we see a systematic
offset between \mdot\ and \lledd. The \mbh\ estimates used here are systematically larger than in Paper I, thus
reducing the values of \lledd[BC] and bringing them more in line with our
estimates of \mdot\ from the thin AD fitting.

The inputs to the Bayesian fitting procedure are \mbh\ and \Mdot, as measured
from the spectra, neither of which require a bolometric correction to calculate.
On the other hand, calculating \lledd\ directly from the spectra requires a
bolometric correction, and the good agreement between \mdot\ and \lledd\ found
here in Fig.~\ref{fig:mdot_comp} indicates that the bolometric correction
factors used in Paper II to calculate \lledd\ give reasonable results.

\subsection{Black Hole Spin}
\label{sec:spin}

\begin{figure*}
  \centering
  \includegraphics[width=140mm]{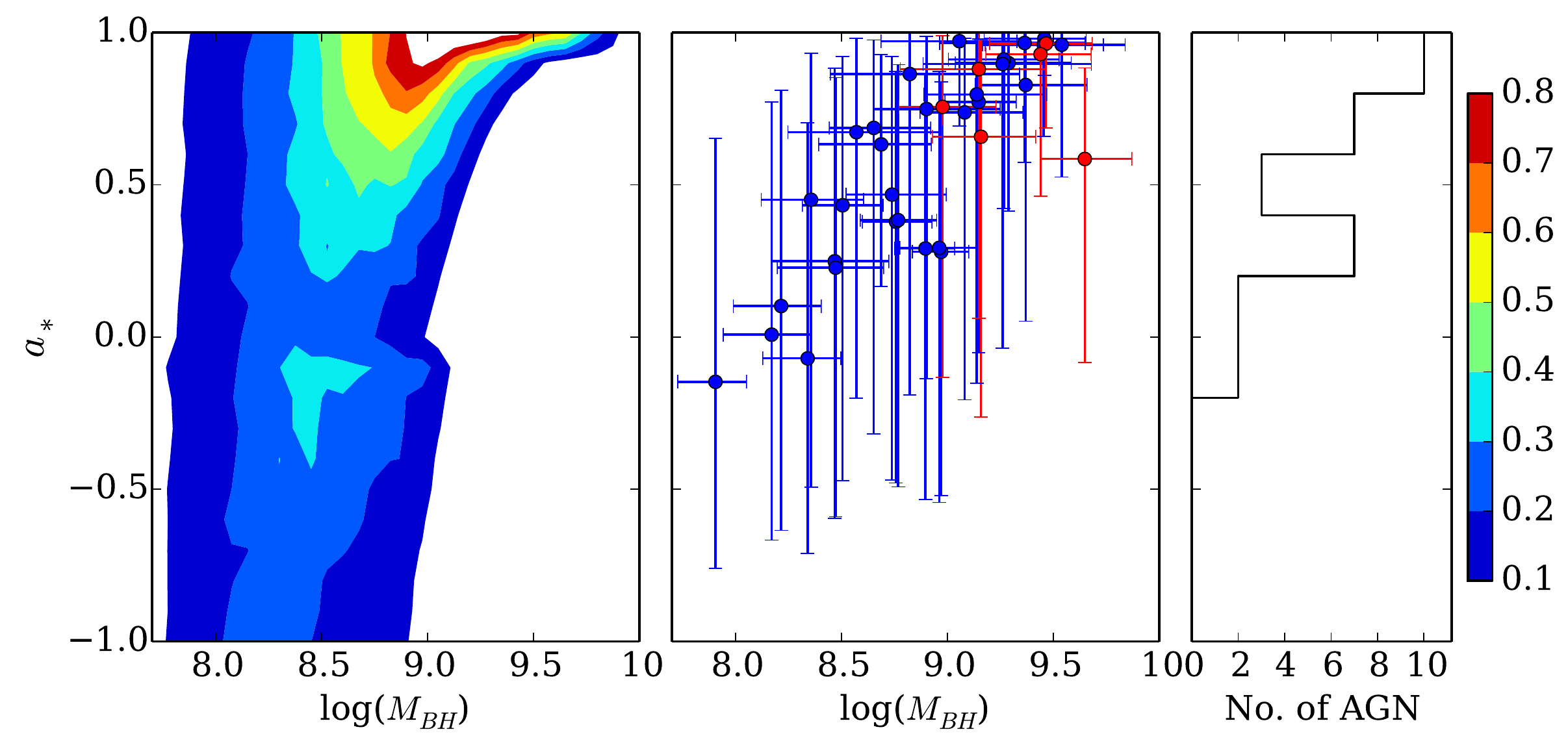}
  \includegraphics[width=140mm]{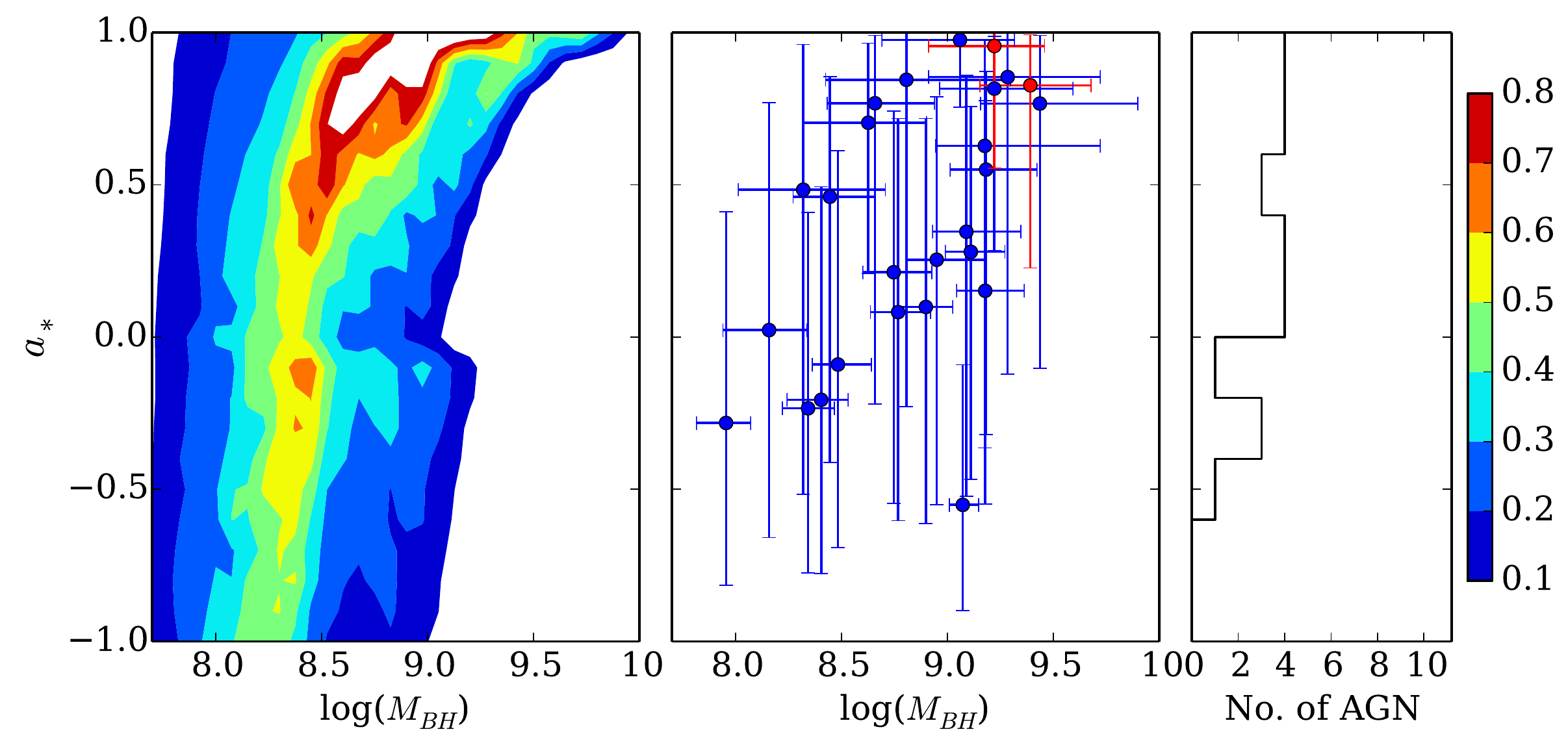}
  \caption{
    The spin parameter, $a_*$, as a function of $M_{BH}$. Top panel is based on
    fits to X-shooter only (37 sources), and the bottom panel is based on fits
    to X-shooter+GALEX (26 sources).
    The left panel is a contour plot of the combined probability distributions
    in $a_*$ and $M_{BH}$ for the sources with satisfactory fits. The middle
    panel shows the median $a_*$ and $M_{BH}$ values, with the red points
    identifying those sources for which dereddening was required for a
    satisfactory thin AD fit. The right panel shows the distribution in the
    best-fit spin parameters.}
   \label{fig:mbh_a}
\end{figure*}

The goal of the spectral fitting is not just to test the thin AD theory, but in
cases where the observed data is consistent with the theory, to attempt to
constrain $a_*$, as demonstrated already in \citetalias{Capellupo15}. With our
results, we see that we can obtain much tighter constraints for active BHs
above $M_{BH}$ $\sim$ 10$^{9}$ \Msun, as compared to those below this mass.
This tendency is expected since precise determination of the spin parameter
depends, crucially, on the wavelength range exhibiting the largest SED
curvature. This range is at longer wavelengths for more massive BHs and BHs
with lower \Ledd. For the most massive objects in our sample, this range is
well inside the X-shooter wavelength coverage, and hence we can better
constrain $a_*$. For lower mass, higher accretion rate BHs, much of the
curvature is at far-UV wavelengths, and the X-shooter range can thus be fitted
by a range of models with a wide range in $a_*$. Fig. \ref{fig:mbh_a} combines
the results presented in Figs \ref{fig:a_mbh_xsh} and \ref{fig:a_mbh_gal} and
Table \ref{tab:master_table}, and it is clear that the most massive BHs have
both the highest spin parameters and the tightest constraints on the spin
parameter.

If we focus on the 17 sources with \mbh\ $>$ 10$^{9}$ \Msun\ and $a_* > 0.7$
(efficiency $\sim$ 0.1), when fitting just the X-shooter spectrum, 10 of those
have a satisfactory fit with GALEX. Of these 10, the estimate of $a_*$ decreases
to below 0.7 for 5 of them after fitting the X-shooter+GALEX SED, and the
errors on $a_*$ are larger. This reduction in spin parameter is due to the
GALEX photometry forcing the fits to lower luminosities at far-UV wavelengths.

We also see that while GALEX provides some crucial information on the SED shape
blueward of $\sim$1200\AA\ for our sample, it does not, in general, reduce the
uncertainties
on the parameters involved in fitting the thin AD model. As mentioned already,
our spin parameter estimates for the highest mass BHs are now more uncertain,
and the uncertainty on the spin for the BHs with \mbh\ $<$ 10$^{9}$ \Msun\ is
similar after including GALEX. This is likely due mostly to the large
uncertainties on the GALEX points. If the `turnover' in the thin AD spectrum
occurs shortward of 1200\AA, then spectra are needed in this
wavelength regime to properly trace the SED and fit the thin AD models.
Follow-up spectroscopy with HST is thus necessary to confidently test the thin
AD model and obtain more precise constraints on the BH spin.

Despite the uncertainties mentioned above, the results still give some insight
into the evolution of SMBH spin in AGN. The two commonly discussed scenarios in
the literature to characterize this evolution are referred to as `spin-up' and
`spin-down'. The difference between these two scenarios is primarily in the
nature of the accretion episodes that fuel the BH. On the one hand, a series of
accretion episodes with random and isotropic orientations will cause the SMBH
to `spin-down' to moderate spins near $a_* \sim 0$, regardless of the final
mass of the SMBH \citep{King08,WangJM09,LiYR12,Dotti13}.
On the other hand, growing a SMBH via a single prolonged accretion episode, or
for the most massive BHs, when the orientations of the accretion episodes have
even a small amount of anisotropy, the SMBH will `spin-up' to a high spin
parameter \citep{Dotti13,Volonteri13}.

In \citetalias{Capellupo15}, we found that our results favour the `spin-up'
scenario, and our current results favour this scenario for similar reasons. We
again find a wide range in spin parameters for the sample,
as shown in the rightmost panels of Fig. \ref{fig:mbh_a}, with the exception
that there are almost no sources with $a_* < -0.5$. Furthermore, even with the
GALEX points included in the analysis, there are many sources with high spin
($a_* > \sim0.5$). If the `spin-down' scenario were dominating, i.e. if there
were multiple, randomly-oriented accretion events throughout the lifetime of
these SMBHs, we would expect a concentration of values around $a_* \sim 0$.
Instead, our results favour scenarios where there is just one long accretion
episode or multiple events with some preferred orientation.

In fact, compared to \citetalias{Capellupo15}, we see a clear shift in the
distribution of $a_*$ towards higher spin. This is due both to the higher black
hole mass estimates (see Section \ref{sec:thinad} and Paper II) and to the
inclusion of an intrinsic reddening correction in the Bayesian fitting
procedure. While for most objects the typical amount of intrinsic reddening is
small ($A_V < 0.15$ mag), any correction of the spectrum for reddening will
increase the luminosity at shorter wavelengths much more than at the longest
wavelengths in the SED. This will favor higher spin parameters, if all other
parameters remain roughly the same.
Previous efforts to constrain BH spin have also generally concluded that
many BHs have high spin, especially the most massive (\mbh\ $> 10^9$ \Msun) BHs
(\citealt{Davis11,Reis14,Reynolds14,Reynolds14a,Trakhtenbrot14}; Wang et al. 2014a).
All of this supports the `spin-up' scenario of BH spin evolution.

\section{Conclusions}

This work is the third in a series of papers describing the spectroscopic
properties of a sample of AGN at $z \sim 1.5$, selected to cover a wide range
in both \mbh\ ($\sim$ 10$^{8}$ to 10$^{10}$ \Msun) and \lledd[BC]
($\sim$0.01 to 0.4) and observed with the X-shooter instrument, which provides
very wide, single-epoch coverage. We apply a similar, but improved,
Bayesian procedure as in Paper I to fit thin AD models to observed AGN SEDs,
this time with
a larger sample (39 AGN),
improved \mbh\ estimates from Paper II,
and the inclusion of intrinsic reddening as a parameter in our Bayesian SED
fitting procedure.

When fitting the thin AD model to the X-shooter spectra alone, we find that we
are able to fit more of the AGN in our sample than in Paper I, with 37 out of
39 AGN (95\%) having a satisfactory fit (Section \ref{sec:xsh_only}).
For those AGN with satisfactory fits, we constrain the spin parameter,
$a_*$, with the constraints becoming less well-defined with decreasing \mbh.
The distribution in $a_*$ for these sources ranges from negative spin to nearly
maximum spin. This distribution tends to
favor the `spin-up' scenario of BH spin evolution, suggesting that these AGN
are generally fueled by relatively long episodes of coherent accretion
with some preferred orientation (Section \ref{sec:spin}).

We also investigate the inclusion of non-simultaneous GALEX photometry in
our analysis.
This decreases the number with satisfactory fits to 26 out of 38 (68\%) sources
(Section \ref{sec:xsh_galex});
however, given the large variability that can occur for AGN
at these UV wavelengths, it is unclear how much variability is affecting our
fitting results for these combined X-shooter+GALEX SEDs.
The inclusion of GALEX photometry also tends to decrease the estimates of
$a_*$, especially for the AGN with larger \mbh, but taken at face value, these
estimates of $a_*$ still support the `spin-up' scenario of BH spin evolution.

While our results support the thin AD theory for a majority of the AGN in our
sample, simultaneous UV and optical spectra are required to properly test the
thin AD theory in the far-UV, where, for many sources, the peak of the thin AD
spectrum occurs. Such simultaneous spectra will also provide the best
constraints on $a_*$, particularly for the lower \mbh\ sources.

\section*{Acknowledgements}

We thank the anonymous referee for helpful comments on the manuscript.
We thank the DFG for support via German Israeli Cooperation grant
STE1869/1-1.GE625/15-1. Funding for this work has also been provided by the
Israel Science Foundation grant number 284/13.
HN acknowledges useful discussions and local support from the International
Space Science Institute (ISSI), in Bern, during a work-group meeting in 2015.



\bibliographystyle{mnras}
\bibliography{full_bibliography_tau}







\bsp	
\label{lastpage}
\end{document}
